\documentclass[aps,prd,nofootinbib,twocolumn,superscriptaddress,preprintnumbers,balancelastpage,longbibliography]{revtex4-1}

\pdfoutput=1
\usepackage{subfigure}
\usepackage{graphicx}
\usepackage{epstopdf}
\usepackage{mathrsfs}
\usepackage{amssymb}
\usepackage{verbatim}
\usepackage{color}
\usepackage{multirow}
\usepackage{amsmath}
\usepackage{hyperref}
\usepackage{physics}

\usepackage{xcolor}

\hypersetup{pdfstartview=FitV,colorlinks=true,linkcolor=blue,citecolor=red,filecolor=black,urlcolor=blue}

\newcommand{\Trh}{T_{\rm RH}}
\newcommand{\Tfo}{T_{\rm FO}}
\newcommand{\Tmax}{T_{\rm max}}
\newcommand{\beq}{\begin{equation}}
\newcommand{\eeq}{\end{equation}}
\newcommand{\bea}{\begin{eqnarray}}
\newcommand{\eea}{\end{eqnarray}}

\newcommand{\rhorh}{\rho_{\rm RH}}
\newcommand{\rhoend}{\rho_{\rm end}}
\newcommand{\arh}{a_{\rm RH}}
\newcommand{\aend}{a_{\rm end}}
\newcommand{\afo}{a_{\rm FO}}
\newcommand{\am}{a_{\rm m}}

\newcommand{\Yfo}{Y_{\rm FO}}
\newcommand{\Yrh}{Y_{\rm RH}}

\newcommand{\noi}{\noindent}

\begin{document}

\preprint{UMN--TH--4422/25, FTPI--MINN--25/04}

\title{Ultra-Relativistic Freeze-Out During Reheating}

\author{Stephen E. Henrich}
    \email[Correspondence email address: ]{henri455@umn.edu}
    \affiliation{William I. Fine Theoretical Physics Institute, School of Physics and Astronomy, University of Minnesota, Minneapolis, Minnesota 55455, USA}

\author{ Mathieu Gross}
\email[Correspondence email address: ]{gross@ijclab.in2p3.fr}
\affiliation{Universit\'e Paris-Saclay, CNRS/IN2P3, IJCLab, 91405 Orsay, France}    

\author{Yann Mambrini}
     \email[Correspondence email address: ]{mambrini@ijclab.in2p3.fr}
\affiliation{Universit\'e Paris-Saclay, CNRS/IN2P3, IJCLab, 91405 Orsay, France}

\author{Keith A. Olive}
    \email[Correspondence email address: ]{olive@umn.edu}
    \affiliation{William I. Fine Theoretical Physics Institute, School of Physics and Astronomy, University of Minnesota, Minneapolis, Minnesota 55455, USA}

\date{May 7th, 2025}

\begin{abstract} 
We perform a thorough investigation of (ultra)relativistic freeze-out (UFO) during reheating. While the standard WIMP (non-relativistic freeze-out) and FIMP (freeze-in) 
paradigms have been explored in detail during the reheating epoch, UFO has 
not been systematically studied, despite the fact that it is operative in a broad region of parameter space. Although dark matter (DM) is ``hot" at the time of relativistic freeze-out, we 
show that it can easily undergo enough cooling by the time of 
structure formation to be compatible with $\Lambda$CDM. Unlike standard WIMP-like freeze-out, there can be significant out-of-equilibrium DM production after UFO, similar to the freeze-in mechanism. However, unlike freeze-in, UFO can accommodate much stronger couplings. The UFO parameter space consistent with $\Omega_{\chi}h^2=0.12$ is quite large, with DM masses spanning about 13 orders of magnitude ($10^{-7} \text{ GeV} \lesssim m_{\chi} \lesssim 10^{6}$ GeV), reheating temperatures spanning 17 orders of magnitude ($10^{-2} \text{ GeV} \lesssim T_{\rm RH} \lesssim 10^{15} \text{ GeV}$)
and Beyond the Standard Model (BSM) effective interaction scales  spanning 11 orders of magnitude ($10^{3} \text{ GeV} \lesssim \Lambda \lesssim 10^{14}\text{ GeV}$). 
Interestingly, the most suitable range of couplings for UFO lies precisely between the typical couplings for WIMPs and FIMPs, rendering UFO quite attractive from the standpoint of detection. Particle physics models that are easily amenable to UFO include heavy vector or scalar portal interactions, along with nonrenormalizable effective interactions. Finally, we show there is a distinction between UV UFO and IR UFO, where the relic abundance for the former is sensitive to the freeze-out temperature, while the abundance for the latter is sensitive to the DM mass and the reheating temperature but insensitive to the freeze-out temperature.
\end{abstract}


\maketitle

\section{Introduction}

The origin, identity and experimental signature of dark matter (DM) are all unknown. 
Of course the three are all inter-related, as the identity of DM would specify its interactions with the Standard Model (SM) and clarify its origin and strategy for detection. 
Concerning the origin of DM in the early universe, a
distinction has been made made between the so-called freeze-out (FO) and 
freeze-in (FI) mechanisms. In the former, the interactions of the DM, although weak, are strong enough to bring it into equilibrium; and the ``WIMP miracle" yields the correct relic abundance thanks to the Boltzmann suppression of the non-relativistic DM number density with respect to the radiation number density while ``freezing out" from the thermal bath. 
In the latter case, DM production occurs continuously but the DM never equilibrates
with SM particles, due to the expansion 
rate of the Universe dominating over the DM production rate. The ``FIMP miracle" is due to the fact that the relic abundance is independent of the DM mass for long-range interactions.

Most studies of the FI/FO production mechanisms have been made in the context
of a radiation domination era \cite{Chu:2011be}, where the evolution of the Universe and the Hubble rate $H$ are dominated by the thermal (SM) plasma. However, radiation domination 
occurs only after a phase called reheating, which generates
the radiation bath. Assuming an early period of exponential expansion driven by the inflaton, the period of reheating begins when inflation ends, though the energy density of the universe is still dominated by the inflaton 
until the inflaton decays and radiation domination begins. If DM is produced during the period of reheating, the 
naive separation between FI and FO becomes much more complex, due to the specific evolution of the temperature, and as a consequence the interaction cross sections, with time.

For example, the naive relations  
$H(T)\sim \sqrt{\rho}\propto T^2$, or $T\propto a^{-1}$, where $a$ represents the cosmological scale factor, do not hold during reheating, while the radiation bath is being created \cite{Scherrer:1984fd,Giudice:2000ex,gkmo1,gkmo2}. Thermalization and the FO mechanism become strongly dependent 
on the evolution of the field dominating the energy density of the 
Universe, and more specifically on the shape of the inflaton 
potential after inflation. Recently, the FIMP scenario has been analyzed in this context \cite{HMO,Bernal:2025fdr}. In addition, non-relativistic WIMP-like freeze-out during reheating has also been considered \cite{Silva-Malpartida:2024emu,Mondal:2025awq}.
However, the possibility of  \textit{ultra-relativistic} freeze-out (UFO) has been unexplored so far. Depending on the strength of the interaction cross section between the DM and SM, 
any one or several of these mechanisms may be responsible for the observed DM abundance. In particular, we will demonstrate here that, for many cross sections and reheating scenarios, UFO constitutes a large intermediate regime that lies between non-relativistic FO and FI. 

One of the central issues concerning the FIMP 
mechanism is the hypothesis that dark matter is not coupled directly to the inflaton, so that its production is secondary, requiring the presence of a thermal bath. 
This can be justified only if the branching ratio for inflaton decays to DM is sufficiently small 
$BR_\phi^\chi \ll 10^{-9} \Trh/m_\phi$, where $\Trh$ is the reheating temperature and $m_\phi$ is the inflaton mass \cite{kmo}. 
Note that the branching ratio cannot be set to arbitrarily small values if the DM is coupled to the SM and the SM is coupled to the inflaton as dark matter will be produced radiatively in inflaton decays or through Bremstrahlung processes \cite{Mambrini:2022uol}. 
Thus, when considering a 
FI mechanism it is necessary to include a more in-depth study of the post-inflationary reheating phase. In clear distinction to the classical non-relativistic FO mechanism where all prior history of the dark matter is erased by entering into thermal 
equilibrium, this is not the case for FI where initial conditions (and interactions) are crucial. It is possible, for example (as we will show in this 
paper), that DM enters into thermal equilibrium {\it 
during} the reheating process and freezes-out relativistically before the end of the 
reheating. This can lead to results that differ significantly from the well-studied case of relativistic freeze-out during radiation domination, which occurs for instance when SM neutrinos freeze-out at temperatures of order 1 MeV.

Though our work is not sensitive to the details of the model of inflation, we will assume that after the period of exponential expansion ends, the inflaton begins a period of oscillations about the minimum of its scalar potential which we parameterize as 
\begin{equation}
    \label{Eq:potmin}
    V(\phi)= \lambda M_P^4 \left(\frac{\phi}{M_{P}}\right)^k, \quad \phi \ll M_{P} \, .
\end{equation}
This occurs for example, in the class of models known as T-models ~\cite{Kallosh:2013hoa} with a potential given by 
\begin{equation}
    V(\phi) \; = \;\lambda M_P^{4}\left|\sqrt{6} \tanh \left(\frac{\phi}{\sqrt{6} M_P}\right)\right|^{k} \, .
\label{Vatt}
\end{equation}
Of primary importance to us here, is the value of $k$
which determines the equation of state parameter $w\equiv P_\phi/\rho_\phi = (k-2)/(k+2)$ during reheating. Here, $P_\phi$ and $\rho_\phi$ are the pressure and energy density of the inflaton condensate oscillating about the minimum of the potential at $\phi = 0$. The overall scale of the potential, determined by $\lambda$, is fixed by the normalization of the cosmic microwave background (CMB) anisotropy spectrum and depends on $k$ \cite{gkmo1}. Note that for $k>4$, the evolution of the inflaton is subject to the effects of fragmentation \cite{Lozanov:2016hid,Garcia:2023dyf} which cause the equation of state to evolve to $k=4$. These effects can be avoided and higher values of $k$ are allowed if the reheating temperature is sufficiently high \cite{Garcia:2023dyf}.  

Assuming some coupling of the inflaton to the Standard Model, the inflaton begins to decay after inflation ends. 
The precise details of the reheating process also depend on the spin statistics of the final state decay products \cite{gkmo2}, which we assume are fermions. 
As the inflaton begins to decay, the radiation
density starting at $\rho_{\rm R} = 0$, rises quickly to some maximum density with temperature $\Tmax$. The density then drops (slowly) as the universe expands, though the inflaton continues to decay. 
The strength of the coupling between the inflaton and SM fermions determines the reheating temperature, $\Trh$, which we take as a free parameter with a lower bound of $\Trh=4$~MeV due to constraints from Big Bang Nucleosynthesis (BBN) \cite{tr4}. 
The end of the reheating period is defined when
the energy density of the inflaton become equal to the energy density in the radiation bath, $\rho_\phi(\arh) = \rho_{\rm R}(\arh)$, where $\arh$ is the scale factor when $T = \Trh$. The relations between $a$ and $T$ will be discussed in more detail below.

Just as we have parameterized the dynamics of the reheating process (with $k, \Trh$, and $\lambda$),
it will be useful to parameterize the interaction cross section between dark matter and Standard Model particles.
 Interactions between DM and SM particles in the early universe can often be characterized by a thermally averaged cross section of the form
\beq
\langle \sigma v \rangle = \frac{T^n}{\Lambda^{n+2}}
\label{anssv}
\eeq 

\noindent where $\Lambda$ is an effective high energy scale associated with a beyond the Standard Model (BSM) interaction, and the value of $n$ will in general depend on the ultra-violet (UV) or infra-red (IR) limit of the interaction.
For example, an interaction between DM scalars and SM scalars (the Higgs doublet) mediated by vector exchange can be described by a cross section
\beq
\langle \sigma v \rangle \sim \frac{g^4 T^2}{(T^2 + M^2)^2} \, ,
\label{anssvbis}
\eeq
where $g^4$ refers to the product of the couplings of the DM and SM fields to the vector mediator, with mass $M$. Examples of interactions that take this form include heavy 
mediators such as $Z'$ portals \cite{Zprime}, motivated by
GUT-type models \cite{SO10}.
At high energies ($T \gg M$), the interaction is similar to that of a contact interaction with $n_h=-2$, whereas at lower energies  ($T \ll M$), $n_l = 2$, with $\Lambda_l \propto M/g$. In this example, when the BSM scale associated with the mediator mass $M$ is below the maximum temperature reached in the thermal 
bath, the dependence of $\langle \sigma v \rangle$ changes from $\propto T^{n_l-4}$ to $T^{n_l}$ for $T<M$.

Eq.~(\ref{anssvbis}) can be further generalized to take into account other types of renormalizable interactions. In addition to the mediator mass, it is also possible that the dark matter coupling is dimensionful (originating from a possible 
derivative coupling).  One way to 
parameterize this dependence is to generalize the cross section as
\beq
\langle \sigma v \rangle \sim \frac{g^4 \mu^{2-n} T^{n}}{(T^2 + M^2)^2} \, ,
\label{Eq:sigmav}
\eeq
where $\mu$ is the dimensionful coupling. 
We then recover the behavior $\langle \sigma v \rangle \sim T^{n-4}/\Lambda^{n-2}$, with $\Lambda \propto \mu$ for $T>M$, and $\langle \sigma v \rangle \sim T^{n}/\Lambda^{n+2}$, with 
\beq \Lambda^{n+2} \propto \mu^{n-2} M^4\,,
\label{Eq:lambdamu}
\eeq
for $T<M$ as in Eq.~(\ref{anssv}). We can interpret this
change of behavior as a change in the power of $T$ from the high temperature regime to the low temperature regime as $n\rightarrow n+4$, or 
$n_l=n_h+4$, and for the scale $\Lambda$ as $\Lambda^{n+2} \rightarrow \Lambda^{n+2}M^4$,
and we will identify $n_l=n$.

Of course the final relic abundance of DM is determined by the interaction (and the expansion history of the Universe), and may
be determined by freeze-out if at some point the DM 
equilibrated with the SM bath, or through freeze-in if 
equilibrium is never attained.  We will take as our condition for equilibrium 
\beq
\Gamma > \frac32 (1+w) H
\label{equilCondition}
\eeq
for an interaction rate $\Gamma$ between the DM and SM.

In what follows, we will systematically examine the conditions for which ultra-relativistic freeze-out plays an important role in the determination of the DM relic abundance.  In Section \ref{sec:rFO}, we first briefly review the notion of UFO as applied during the radiation dominated epoch, using SM neutrinos as an example. In Section \ref{sec:rFOrh}, we consider  UFO during reheating. In Section \ref{sec:rFOcond}, we first derive the conditions for
UFO and then calculate the freeze-out temperature in Section \ref{sec:rFOtemp}. From the conditions, we are able to set limits on the reheating temperature for this mechanism to be viable. These limits are derived in Section \ref{sec:rFOlim}. Then in Section \ref{sec:DMden}, we compute the number density and relic abundance of DM from ultra-relativistic freeze-out.
Our conclusions are given in Section \ref{sec:concl}.

\section{Relativistic Freeze-Out in the Radiation Dominated Era}
\label{sec:rFO}

It is instructive to first review (ultra)relativistic freeze-out during a standard radiation dominated era, assuming instantaneous reheating. We will then compare these results to those obtained with non-instantaneous reheating. 

The classic example of relativistic freeze-out is that of the light (left-handed) neutrinos in the SM. In this section, we consider DM that freezes out relativistically during the radiation dominated era, in a similar manner to neutrinos. In this case, $m_\chi \ll \Tfo \ll M$, where $m_\chi$ is the DM mass and $M$ is the electroweak scale. For neutrinos, we may take $m_\chi = m_\nu$, or even $\Sigma m_\nu$, and the mediator mass is related to the weak-scale gauge bosons. 
The relativistic freeze-out temperature during a radiation dominated era (with $w = 1/3$) can be found by equating the interaction rate to the expansion rate as follows:
\beq
\Gamma(\Tfo) = 2 H(\Tfo)
\label{Equil:relFOrad}
\,,
\eeq
or
\beq
\langle \sigma v \rangle(\Tfo) \,n_\chi(\Tfo) = 2 \frac{\sqrt{\rho_{R}(\Tfo)}}{\sqrt{3} M_{P}}\,,
\label{Eq:focondition}
\eeq
where $\rho_R$ is the energy density stored in the thermal bath, 
given by 
\beq
\rho_R(T)=\alpha T^4\,,
\eeq
and $\alpha = \frac{\pi^2}{30}g_{*}(T)$, $g_{*}(T)$ being the number of relativistic degrees of freedom,
$g_* = 427/4$
in the Standard Model radiation bath for large $T$. Importantly, Eq. (\ref{Equil:relFOrad}) should be imposed with the correct form of $\Gamma(T)$ in the relativistic regime, where the equilibrium number density of DM is $n_{\chi} \propto T^3$. In Eq.~(\ref{Eq:focondition}), $\langle \sigma v \rangle$ is the interaction cross section and $M_P \simeq 2.4 \times 10^{18}$~GeV 
is the reduced Planck mass. With the ansatz (\ref{anssv}), and supposing $g_\chi$ internal degrees of freedom for the dark matter 
$\chi$, Eq.~(\ref{Eq:focondition}) then becomes 
\beq
\frac{g_\chi \zeta(3)}{\pi^2} \frac{\Tfo^{n+3}}{\Lambda^{n+2}} = 2 \sqrt{\frac{\alpha}{3}}\frac{\Tfo^2}{M_{P}}\,,
\eeq
or
\beq
\Tfo = \Lambda\, \left(\frac{\Lambda}{M_{P}}\right)^{\frac{1}{n+1}}\left(\frac{2 \pi^2}{g_\chi \zeta(3)} \sqrt{\frac{\alpha}{3}}\right)^{\frac{1}{n+1}}\,.
\label{Eq:TFO}
\eeq
where
\[g_\chi = \begin{cases} 1 \text{, for real scalars} \\ \frac{3}{2} \text{, for Majorana fermions.}\end{cases} \]
For weak scale interactions, and $\Lambda \sim 100$~GeV, $\Tfo \sim 1$~MeV for $n=2$.

The DM number density at relativistic freeze-out is given by 
\beq
n_{\chi}(\Tfo) = \frac{g_\chi \zeta(3)}{\pi^2}\Tfo^3 \, .
\label{ntfo}
\eeq
To obtain the present-day relic abundance of dark matter, the density in Eq.~(\ref{ntfo}) must be redshifted to the present
by 
\beq
\left(\frac{a_{\rm FO}}{a_0}\right)^3 = \left( \frac{T_0}{\Tfo}\right)^3 \frac{g_0}{g_{\rm FO}} \, ,
\label{today}
\eeq
where $T_0$ is the present temperature of the CMB, 
$g_0 = (43/4)(4/11)$,  $g_{\rm FO}$ is the number of degrees of freedom at the freeze-out temperature, $\Tfo$, and $a_i$ is the corresponding 
scale factor at $T_i$\footnote{We are assuming that $\Tfo \gtrsim 1$~MeV, to avoid strong constraints on $N_{\rm eff}$ from big bang nucleosynthesis (BBN) \cite{ysof}. Indeed, even adding a single new real scalar or Majorana fermion would require a freeze-out temperature in excess of the QCD confinement transition.}.
We can then find the fraction of critical density today compared to the DM density determined by Planck \cite{Planck:2018vyg}
\beq
\frac{\Omega_\chi h^2}{0.12} 
\simeq g_\chi 
\left(\frac{106.75}{g_{\rm FO}}\right)
\left( \frac{m_\chi}{{\rm 170~{\rm eV}}}\right)\,.
\label{Oh2rfo}
\eeq
Importantly, we see that the relic density does not explicitly dependent upon $\Lambda$, $T_{\rm RH}$, or $\Tfo$, 
but exhibits indirect dependence on the latter only through ${g_{\rm FO}}$. Thus, there is a narrow 
mass range consistent with the observed relic density which 
depends exclusively on
$g_{\rm FO}$ and $g_\chi$. Requiring $\Omega_{\chi} h^{2} < 0.12$ for relativistic freeze-out after reheating thus leads to the following constraints on $m_{\chi}$:
\begin{align}
m_{\chi} & < 17  \text{ eV} \qquad g_{\rm FO} = 43/4 \\
m_{\chi} & < 170  \text{ eV} \qquad g_{\rm FO} = 427/4
\label{Eq:masslimithot}
\end{align} 
for (real) scalar DM and 
\begin{align}
m_{\chi} & < 11 \text{ eV} \qquad g_{\rm FO} = 43/4 
\label{2cfl} \\
m_{\chi} & < 110  \text{ eV} \qquad g_{\rm FO} = 427/4
\end{align} 
for Majorana fermion DM, depending on the 
value of $g_{\rm FO}$ at decoupling time. 
These are modern versions of the bounds found in 1966 by Gershtein and Zeldovich \cite{Gershtein:1966gg}
(also later by Cowsik and McClelland\footnote{Note that, contrarily to the Zeldovich paper, the Cowsik McClelland forgot to take into account the $g_s^i$ factors in their calculation.} \cite{Cowsik:1972gh} and Szalay and Marx \cite{Szalay:1974jta}) corresponding to the maximum neutrino mass such that the universe will not overclose (i.e. they were interested in $\Omega = 1$ rather than $\Omega h^2 = 0.12$). For neutrinos, with $g^{\rm FO} = 43/4$, and $m_\chi \to \Sigma m_\nu$, Eq.~(\ref{Oh2rfo}) becomes 
\beq
\frac{\Omega_\chi h^2}{0.12} 
\simeq 
\left( \frac{\Sigma m_\nu}{{\rm 11~{\rm eV}}}\right)\,,
\label{Oh2nu}
\eeq
for two component neutrinos, equivalent to Eq.~(\ref{2cfl}).

Therefore, in order for a  DM candidate to freeze out relativistically 
during the radiation dominated era after reheating
and account for the observed relic density, we require $ 11 \text{ eV} \lesssim m_{\chi} \lesssim 170 \text{ eV}$. 
These 
limits are independent of $\Lambda$, $T_{\rm RH}$, and $\Tfo$, provided that relativistic freeze-out occurs for the 
relevant interaction. With DM masses on this scale, it would be 
challenging to construct a model that circumvents the well-known 
problems associated with hot DM like structure formation
or Lyman-$\alpha$ constraints, which 
is one reason why this mechanism (UFO) has 
not been extensively studied outside the context of neutrinos. Indeed, constraints from the shape of the matter power spectrum combined with CMB data yield very strong constraints on the sum of neutrino masses, 
$\Sigma m_\nu \lesssim 0.1$~eV \cite{Planck:2018vyg,Garny:2020rom}.
However, we will see that the situation is strikingly different 
when UFO occurs during reheating 
({\it i.e.} when we do not assume instantaneous reheating).

\section{Ultra-Relativistic Freeze-Out During Reheating}
\label{sec:rFOrh}

The reheating period begins when the period of exponential expansion ends, and continues until the time when the Universe is dominated by radiation. During reheating, the energy density of the Universe is dominated by 
the inflaton condensate (or free inflaton quanta) until inflaton decay is complete. Early decays begin to populate the Universe with SM fields reaching a maximum temperature which
redshifts in a model dependent way during the reheating process.
Thus we have the possibility that the DM will come into \textit{and} go out of equilibrium during the reheating process. The latter may occur when the DM is non-relativistic (as is common in most freeze-out scenarios), or alternatively when the DM is still relativistic.
This is the central topic of our discussion. 

In the sections to follow, we consider a real scalar dark matter candidate for specificity ($g_{\chi}=1$). Scalar fields coming out of an inflationary epoch can pick up large scale fluctuations which can lead to a large dark matter abundance independent of any freeze-in or freeze-out scenario \cite{Turner:1987vd, Peebles:1999fz, Enqvist:2014zqa, Markkanen:2018gcw, Cosme:2020nac,Lebedev:2022cic,Choi:2024bdn,Garcia:2025rut}. These may be suppressed if there are large self interactions for the scalar. For example, if $V(\chi) = \frac12 m_\chi^2 \chi^2 + \lambda_\chi^4$, the contribution to the relic density from fluctuations is
$\Omega_\chi h^2/0.12 = 2 \times 10^{-17} \lambda_\chi^{-\frac34} m_\chi \Trh$~GeV$^{-2}$.
This will be negligible for $\lambda_\chi \sim 1$ in the cases considered below. Note that while we use $g_{\chi}=1$ in our calculations below, fermionic DM is also perfectly compatible with UFO during reheating.

\subsection{Conditions for ultra-relativistic freeze-out}
\label{sec:rFOcond}

Considering the possibility of ultra-relativistic freeze-out 
{\it during} reheating, we first notice that there are some interaction types (i.e. some values of $n$) for which 
UFO won't be possible. This is because UFO requires not only that the interaction 
allows the dark component to
reach equilibrium at high temperatures, but also that the 
temperature dependence of the interaction $\Gamma$ is steeper 
than the temperature dependence of $H$, to ensure the possibility that $H(\Tfo)=\Gamma(\Tfo)$ with $H>\Gamma$ for $T<\Tfo$. Thus, we can determine a 
critical, minimum value of $n$ such that UFO becomes 
possible. In particular, for $\Gamma(T) \propto T^{\gamma_1}$ in 
the relativistic regime and $H(T) \propto T^{\gamma_2}$, we 
require $\gamma_{1} > \gamma_{2}$ for UFO to 
be possible, where the interaction rate in the relativistic regime is 
given by~\footnote{Note that for all the interactions considered, UFO is not possible when $T>M$ as $\gamma_1 < \gamma_2$ for all $k$ in this case. Thus we use the low $T$ form of the cross section as in Eq.~(\ref{anssv}). }
\begin{equation}
\Gamma(T) = \langle \sigma v \rangle n = \frac{T^{n}}{\Lambda^{n+2}} \frac{g_\chi \zeta(3)}{\pi^2} T^3\,.
\label{Eq:gamma}
\end{equation} 
To compare the temperature-dependence of $\Gamma(T)$ to $H(T)$, we must 
first evaluate $H(T)$ during the reheating period. As discussed earlier, during reheating, the energy budget of the universe is dominated by the inflaton field $\phi$. 

From the standard Boltzmann equations of motion for $\rho_\phi$, 
and $\rho_R$, one deduces for $a \ll \arh$ \cite{gkmo1,gkmo2}
\beq
\rho_\phi(a)=\rhorh\left(\frac{\arh}{a}\right)^\frac{6k}{k+2}\,,
\label{Eq:rhophi}
\eeq
and
\beq
\rho_R\simeq
\begin{cases}
\rhorh\left(\frac{\arh}{a}\right)^{\frac{6k-6}{k+2}}~~~~k<7\,,
\label{Eq:rhoR}
\\
\rhorh\left(\frac{\arh}{a}\right)^4~~~~~~~~k>7
\,,
\end{cases}
\eeq
where we assumed reheating due to inflaton decay with a fermionic final state. Note that for $a < \arh$, the dependence of $\rho_{tot}\simeq \rho_\phi$ on $T$ is always steeper 
($\propto T^{\frac{4 k}{k-1}}$ and $\propto T^{\frac{6k}{k+2}}$
for $k<7$ and $k>7$ respectively) than in the radiation domination era ($a > \arh$)
where $\rho_{tot}\simeq \rho_R$.  The smallest slopes are indeed $\rho_\phi\propto T^{\frac{24}{5}}$ for $k=6$ and $k=8$.
We then expect freeze out to occur 
earlier (smaller $a$) for non-instantaneous reheating.

Eq.~(\ref{Eq:rhoR}) is only valid for $a \gg \aend$, where $\aend$ is the scale factor at the end of inflation. Initially, $\rho_{\rm R}(\aend) = 0$,
and climbs rapidly to a maximum at $a = a_{\rm max}$, before following the power-law behavior in Eq.~(\ref{Eq:rhoR}). 
We define 
\beq
\beta_k = \left(\frac{3k-3}{2k+4}\right) \text{, and } \gamma_k = \left(\frac{3k-3}{7-k}\right).
\eeq
The ratio of $\aend$ to $a_{\rm max}$, the scale factor when $T = \Tmax$, is given by \cite{gkmo2}
\beq
\frac{a_{\rm end}}{a_{\rm max}}=\beta_{k}^{\frac{k+2}{14-2k}} \, .
\eeq
Since we treat $\Trh$ as a free parameter, we can express the temperature, $\Tmax$ as a function of $\Trh$,
\beq
\alpha \Tmax^4 = \left( \alpha \Trh^4 \right)^\frac1k \rhoend^\frac{k-1}{k} \gamma_k^{-1} \beta_k^\frac{2k+4}{7-k} \, ,
\label{tmax}
\eeq
for $k<7$, and
\beq
\alpha \Tmax^4 = \left( \alpha \Trh^4 \right)^\frac{k-4}{3k} \rhoend^\frac{2k+4}{3k} \gamma_k^{-1} \beta_k^\frac{2k+4}{7-k} \, ,
\label{tmax2}
\eeq
for $k>7$, where $\rhoend$ is the inflaton energy density at the end of inflation.

Combining Eqs.~(\ref{Eq:rhophi}) and (\ref{Eq:rhoR}), the Hubble parameter during reheating for $k<7$ is 
\begin{equation}
H(T) = \frac{\sqrt{\rho_{\phi}}}{\sqrt{3} M_{P}} = \sqrt{\frac{\alpha}{3}} \frac{T_{\rm RH}^2}{M_{P}} \left(\frac{T}{T_{\rm RH}}\right)^{\frac{2k}{k-1}}\,,
\label{Eq:HT}
\end{equation} 
and
\beq
H(T) = \sqrt{\frac{\alpha}{3}} \frac{T_{\rm RH}^2}{M_{P}} \left(\frac{T}{T_{\rm RH}}\right)^{\frac{3k}{k+2}}\,,
\label{Eq:HTkgeq7}
\eeq
for $k>7$.
Hence, $\gamma_{1} = n+3$ and for $k<7$, $\gamma_{2} = \frac{2k}{k-1}$. We therefore find the critical value of $n$ above which UFO during reheating becomes possible to be 
\begin{equation} 
n>n_{c}^{k<7} = \frac{3-k}{k-1}\,.
\label{Eq:infnkl7}
\end{equation}
For $k>7$, $\gamma_2 = 3k/(k+2)$, and the condition  $\gamma_1 > \gamma_2$ becomes
\beq
n>n_c^{k>7}=-\frac{6}{k+2}\,.
\label{Eq:infnkm7}
\eeq

  For $k=2$, $4$, and $6$,  $n_{c} = 1$, $-1/3$, and $-3/5$ respectively. An immediate consequence of this, 
is that UFO during 
reheating with $k<7$ is never possible 
for $n= -2$. This corresponds, for example, to interactions between DM scalars and SM scalars via a scalar mediator (SSS), or to a simple contact interaction, 
where the scattering is highly efficient at low temperature. In these cases, the only two possibilities are nonrelativistic freeze-out (e.g. the WIMP scenario) or freeze-in (i.e. the FIMP scenario). Thus, if the DM reaches equilibrium for $n < n_{c}$, it will stay in thermal equilibrium with the Standard Model down to a temperature $ T = \Tfo < m_\chi$, when the annihilation rate exceeds the production rate as in the classical non-relativistic freeze out case. More generally, Eqs.~(\ref{Eq:infnkl7}) and (\ref{Eq:infnkm7}) exclude long range interactions with cross sections which are $\propto 1/T^2$ from ever exhibiting UFO. Note that long-range interactions also do not allow for UFO in the radiation--dominated epoch, where it is easy to calculate that $n_{c} = -1$.

In contrast, for interactions between fermionic DM and SM fermions with a heavy 
(scalar or vector) mediator (FVF or FSF) where $n=2$, UFO 
during reheating will always be possible for a broad range of $\Lambda$, $T_{\rm RH}$, and $m_{\chi}$ values (as long as $m_{\chi} \ll M$, where $M$ is the mediator mass). Thus, as a first conclusion, 
UFO with an interaction cross section 
$\langle \sigma v \rangle \propto T^n$, with $n > -1$ is possible during reheating, as well as after reheating.

In addition to the necessary condition, that $\gamma_1 > \gamma_2$
to allow
for UFO, we must also ensure that the DM comes into thermal equilibrium some time during
reheating.  
For $M > T_{\rm max}$, this minimal requirement for equilibrium is $\Gamma(T_{\rm max}) > \frac{3k}{k+2} H(T_{\rm max}) \sim \frac{3k}{k+2} H_{\rm end}$. 
Indeed, because the rate increases as $T$ increases from $0$ to $\Tmax$, before 
decreasing as $T^{\gamma_1}$, the maximal rate and cross section is achieved at $\Tmax$. If $\Gamma(\Tmax)< \frac{3k}{k+2} H(\Tmax)$, because one needs
$\gamma_1>\gamma_2$
to ensure a successful freeze-out, $\chi$ would never enter in  thermal equilibrium after $\Tmax$.
Alternatively, if $\Tmax>M$, UFO may still occur if $\Gamma(\Tmax)< \frac{3k}{k+2} H(\Tmax)$; but to ensure thermalization, for $\Tmax>T>M$, we 
have an additional lower bound on $\gamma_2$ so that
\begin{align}
n  - 1 &  < \frac{2k}{k-1}  < n+3 \qquad k<7 \,, 
\label{Eq:supnkl7}
\\
n - 1  &  < \frac{3k}{k+2} < n+3 \qquad k>7
\label{Eq:supnkm7}\, ,
\end{align}
where the first inequality is needed to ensure thermalization, and the second inequality is needed to ensure relativistic freeze-out as in Eqs.~(\ref{Eq:infnkl7}) and (\ref{Eq:infnkm7}).
These conditions are satisfied for all of the renormalizable examples considered here\footnote{For non-renormalizable interactions, the upper limits on $n$ given by Eqs.~(\ref{Eq:supnkl7}) and (\ref{Eq:supnkm7}) do not apply, since the origin of these limits is tied to the temperature dependence of the interaction switching above a heavy mediator. As a result, $n$ can take any value for non-renormalizable interactions and still be consistent with UFO, as long as $\Gamma(T_{\rm max}) > H(T_{\rm max})$. Note that the lower limits of $n>n_{c}$ given by Eqs.~(\ref{Eq:infnkl7}) and (\ref{Eq:infnkm7}) are automatically satisfied for non-renormalizable interactions.}.
From Eq.~(\ref{Eq:supnkl7}) for $k<7$, and Eq.~(\ref{Eq:supnkm7})
for $k>7$, we obtain for $\Tmax>M$ a range of allowed values in the ($n$, $k$) parameter space. This gives for instance, 
$1<n<5$ ($-\frac34<n<\frac{14}{4}$) for $k=2$ ($k=6$). Note that the upper bound is 4 units more than the lower bound, which is
easy to understand as the cross section takes on 4 units (in the exponent) between $T>M$
and $T<M$.

To summarize, the conditions (\ref{Eq:infnkl7}) and (\ref{Eq:infnkm7})
combined with $\Gamma(\Tmax)> \frac{3k}{k+2} H(\Tmax)$ are {\it necessary}
and {\it sufficient} conditions to ensure UFO before $\Trh$ if $M>\Tmax$. For $\Tmax>M$, 
one needs to add the conditions (\ref{Eq:supnkl7}) and (\ref{Eq:supnkm7})
combined with $\Gamma(M)> \frac{3k}{k+2} H(M)$. These constraints can be translated into constraints on $\Trh$, which we discuss in Section C. Because there will be additional constraints on $\Trh$ that are related to the freeze-out temperature, $\Tfo$, we now turn to determining $\Tfo$.

\subsection{Determining the relativistic freeze-out temperature during reheating}
\label{sec:rFOtemp}

We can readily determine the freeze-out temperature, $\Tfo$, as a function of $n$, $k$, $\Lambda$, and $\Trh$ using our condition for equilibrium, namely Eq. (\ref{equilCondition}). For UFO, we must ensure that both the DM number density and the thermally averaged cross section are taken in the relativistic regime, with 
$M > T > m_\chi$. To do this, we simply substitute Eq. (\ref{Eq:gamma}) into Eq. (\ref{equilCondition}) and make the inequality in Eq. (\ref{equilCondition}) an equality. For a particular interaction, we then determine the temperature at which UFO during reheating will occur (if it does in fact, occur) from
$\Gamma(\Tfo) = \frac{3k}{k+2}H(\Tfo)$, and requiring 
\begin{equation}
    M > \Tfo> T_{\rm RH} \text{, and  } m_{\chi} < \Tfo.
    \label{tfoConditions}
\end{equation}
If there is no $\Tfo$ that satisfies the above conditions, then  either 1) freeze-out will not occur (i.e. the interaction never reached equilibrium with the Standard Model radiation bath), 2) non-relativistic freeze-out may occur prior to reheating, or 3) freeze-out (relativistic or non-relativistic) may occur after reheating during the radiation-dominated era. We already covered the case of relativistic freeze-out during radiation domination in Section \ref{sec:rFO}. If the inequalities in (\ref{tfoConditions}) do hold, then UFO during reheating will occur and the freeze-out temperature is given by (for $k < 7$)
\beq
\Tfo^{k<7}=\left(\frac{3k}{k+2}\frac{\pi^2}{g_{\chi}\zeta(3)}\sqrt{\frac{\alpha}{3}}\frac{\Lambda^{n+2}}{M_P} \right)^\frac{k-1}{k+nk-n-3}\Trh^{\frac{2}{n+3-k-kn}}\,,
\label{Eq:tfokl7}
\eeq
which gives for $k=2$, $n=2$, and $g_\chi = 1$, 
\begin{align}
\Tfo^{k=2}|_{n=2} & = \sqrt{\frac{\alpha}{3}}\frac{3\pi^2}{2g_{\chi} \zeta(3)} \frac{M^4}{g^4 T_{\rm RH}^2 M_{P}} \nonumber
\\
&
\simeq 1.7\times 10^9\left(\frac{M/g}{10^{10}~\rm GeV}\right)^4
\left(\frac{10^7~\rm GeV}{\Trh}\right)^2~\rm GeV
\,,
\label{Eq:tfok2n2}
\end{align}
where we used $\Lambda=M/g$, from Eq.~(\ref{Eq:sigmav}).

For $k>7$, we obtain
\beq
\Tfo^{k>7}=\left(\frac{3k}{k+2}\frac{\pi^2}{g_\chi \zeta(3)}\sqrt{\frac{\alpha}{3}}\frac{\Lambda^{n+2}}{M_P}\right)^{\frac{k+2}{6+2n+nk}}\Trh^{\frac{4-k}{6+2n+nk}}\,,
\label{Eq:tfokm7}
\eeq
which gives for $k=8$, $n=2$, and $g_\chi = 1$,
\beq
\Tfo^{k=8}|_{n=2}\simeq 8.7\times 10^7\left(\frac{M/g}{10^{10}~\rm GeV}\right)^\frac{20}{13}\left(\frac{10^7~\rm GeV}{\Trh}\right)^\frac{2}{13}~\rm GeV\,,
\eeq
for a real scalar DM candidate.

Note that this treatment of freeze-out neglects any strong resonant production. However, even in the case of strong resonant production, our estimates above will cover all cases except where the interaction \textit{only} comes into equilibrium for a brief period near the resonance peak. In this latter case, the abundance is easy to estimate. For example, in the case of a heavy $Z'$ mediator, large production near the pole may briefly bring the interaction into equilibrium. The freeze out temperature for such cases will be very near the value of $M_{Z'}$ itself and we can simply evolve the DM number density after freeze-out using the Boltzmann equation for out-of-equilibrium dynamics.

\subsection{Constraints on $\Trh$}
\label{sec:rFOlim}

The preceding discussion in Sections IIIA and IIIB have now prepared us to obtain and analyze a collection of non-trivial constraints on $\Trh$ that are required for UFO during reheating. The origin of these constraints can be summarized as follows: 1) We require that the dark component was in equilibrium with the SM radiation bath at early times (high temperatures); 2) We require that $\Tfo > m_{\chi}$, since otherwise freeze-out will be semi-relativistic or non-relativistic; and 3) We require that $\Tfo > \Trh$ since otherwise we have freeze-out during the radiation-dominated era as already discussed in Section \ref{sec:rFO}. In this section, we will translate each of these conditions into constraints on $\Trh$ for fixed values of $n$, $k$, and $\Lambda$. 

\subsubsection{Minimal conditions for equilibrium}

We first consider the condition that the dark component must be in equilibrium at early times. As discussed in Section \ref{sec:rFOcond}, the minimal condition for equilibrium to be reached at some time during reheating is $\Gamma(\Tmax)>\frac{3k}{k+2}H(\Tmax)$ for $M > T_{\rm max}$ or $\Gamma(M)> \frac{3k}{k+2} H(M)$ for $M < T_{\rm max}$ . These constraints can be translated to minimal values of the reheating temperature required for equilibrium. If $T_{\rm RH}$ is less than these minimum values, the interaction will never reach equilibrium and the DM relic abundance will be determined by FI. These constraints, separated by $M > T_{\rm max}$ or $M < T_{\rm max}$ and $k<7$ or $k>7$ are reported below.

\noi i) For $M > \Tmax$ and $k<7$, 
\beq
\Gamma(\Tmax)>\frac{3k}{k+2}H(\Tmax) \Rightarrow
\nonumber
\eeq
\begin{align}
\Trh&> \beta_{k}^{\frac{(nk+2n+6)k}{(2k-14)(n+3)}}\gamma_{k}^{\frac{k}{4}}\alpha^{\frac{k-1}{4}} \nonumber \\ & \times \left(\frac{3k}{k+2}\frac{\pi^2}{g_{\chi}\zeta(3)}\frac{\Lambda^{n+2}}{\sqrt{3} M_P}\right)^\frac{k}{n+3}\rhoend^{\frac{n+3-k-nk}{4(n+3)}} \,,
\label{Eq:trhmintmaxkl72}
\end{align}
which for $k=2$, $n=2$, and $g_\chi =1$ gives 
\beq
\Trh> 2.3 \times 10^6\left(\frac{\Lambda}{10^{10}~\rm GeV}\right)^\frac85~\rm GeV\,,
\eeq
where we took $\rhoend^\frac14=5.2\times 10^{15}$ GeV (see e.g., \cite{Garcia:2025rut}).

\noi ii) For $M > \Tmax$ and $k>7$,  we obtain 
\begin{align}
\Trh&> (-\gamma_{k})^{\frac{3k}{4(k-4)}}\beta_{k}^{\frac{3k(nk+2n+6)}{(2k-14)(k-4)(n+3)}} \alpha^{\frac{k+2}{2k-8}} \nonumber \\ & \times \left(\frac{3k}{k+2}\frac{\pi^2}{g_{\chi}\zeta(3)}\frac{\Lambda^{n+2}}{\sqrt{3}M_P}\right)^\frac{3k}{(k-4)(n+3)}
\rhoend^{\frac{-nk-2n-6}{2(k-4)(n+3)}}.
\label{Eq:trhmintmaxkg72}
\end{align}

Next, we consider $M < \Tmax$. Combining Eqs.~(\ref{Eq:sigmav}) and (\ref{Eq:HT}), we obtain
from $\Gamma(M)>\frac{3k}{k+2} H(M)$ (or equivalently, requiring $\Tfo < M$)  for $k<7$
\beq
\frac{g_\chi \zeta(3)}{\pi^2}\frac{M^{n+3}}{ \Lambda^{n+2}}> \frac{3k}{k+2} \sqrt{\frac{\alpha}{3}}
\frac{\Trh^2}{M_P}\left(\frac{M}{\Trh}\right)^\frac{2k}{k-1}\,,
\eeq
and this can be translated into a lower bound on $\Trh$.

\noi iii) For $M < \Tmax$ and $k<7$, 
\beq
\Trh > \left(
\frac{3 k \pi^2}{g_\chi \zeta(3) (k+2)}
\sqrt{\frac{\alpha}{3}}
\frac{\Lambda^{n+2}}{M_P}
\right)^\frac{k-1}{2}
M^\frac{n+3-k-nk}{2}\,,
\label{Eq:trhmintmaxkl72g}
\eeq
 where 
\beq
\Lambda=\mu^{\frac{n-2}{n+2}}\left(\frac{M}{g}\right)^{\frac{4}{n+2}}\,.
\label{Eq:lambdam}
\eeq
For $k=2$, $n=2$, $g=1$, and a real scalar DM candidate ($g_\chi=1$), we find
$\Lambda=M/g$ (from Eq.~(\ref{Eq:lambdam})), and
\beq
\Trh>4.2\times 10^6~\left(\frac{\Lambda}{10^{10}~\rm GeV}\right)^\frac32~{\rm GeV}\,.
\label{mlttmax22}
\eeq

\noi iv) For $M < \Tmax$ and $k>7$, 
\beq
\Trh> \left(
\frac{3 k \pi^2}{g_\chi \zeta(3) (k+2)}
\sqrt{\frac{\alpha}{3}}
\frac{\Lambda^{n+2}}{M_P}
\right)^\frac{k+2}{k-4}
M^\frac{-6-2n-nk}{k-4}\,.
\label{Eq:trhmintmaxkg72g}
\eeq
Note that the lower limits in cases iii) and iv) were obtained by taking $\Tfo = M$.

\begin{figure}[!ht]
\centering
\vskip .2in
\includegraphics[width=3.3in]{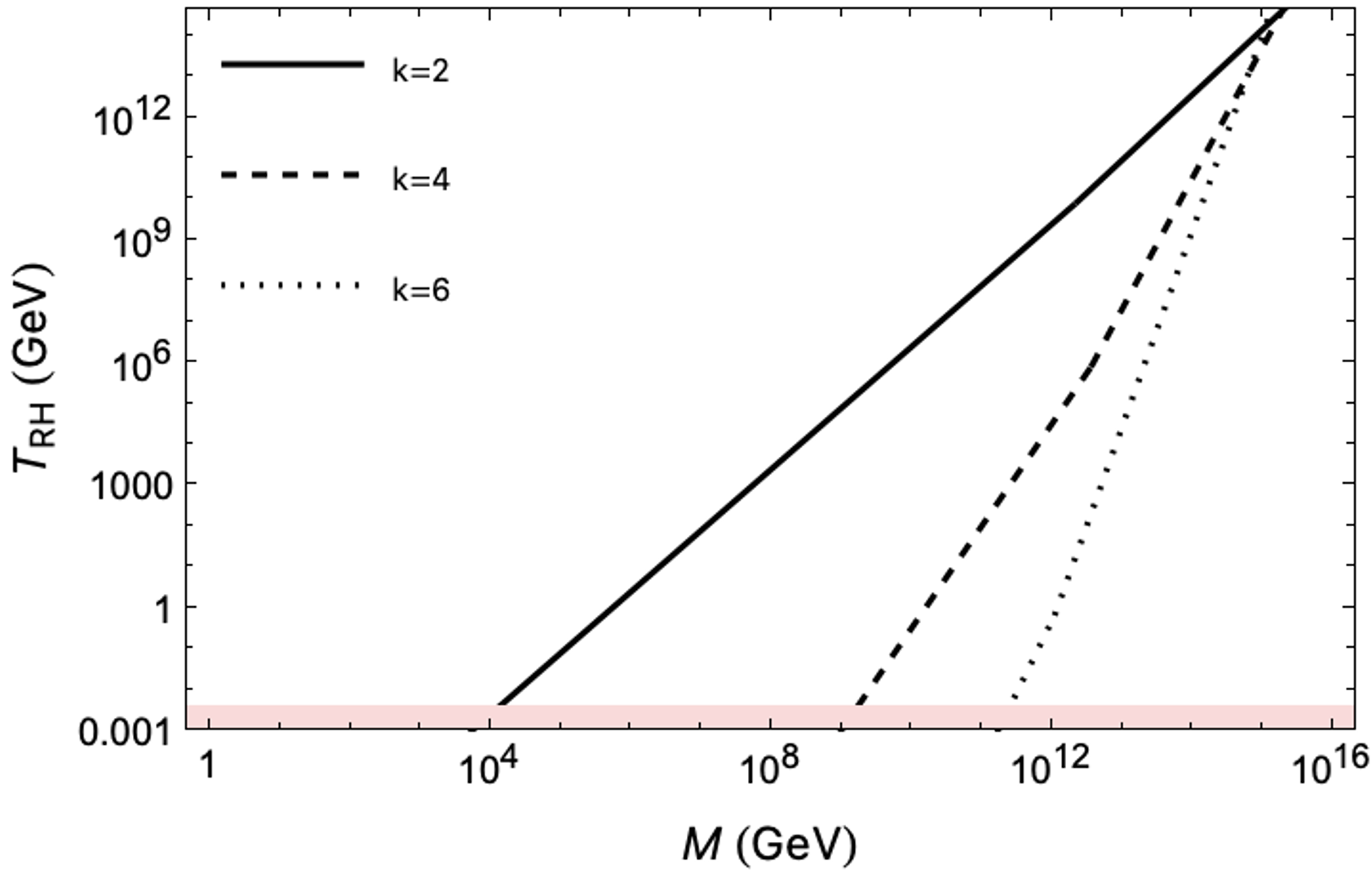}
\caption{\em \small The lower bound on $\Trh$ vs. $M$ for $k=2$, $k=4$, and $k=6$, and $n=2$ with $g=1$ so that $M=\Lambda$. Regions above each line correspond to the interaction reaching equilibrium during the reheating epoch. As a result, relativistic or non-relativistic freeze-out become possible in these regions. The red shaded region corresponds reheating temperatures below the BBN limit of $T_{\rm RH}=4$ MeV \cite{tr4}.
}
\label{trhCritvsLambda}
\end{figure}

\begin{figure}[!ht]
\centering
\vskip .2in
\includegraphics[width=3.3in]{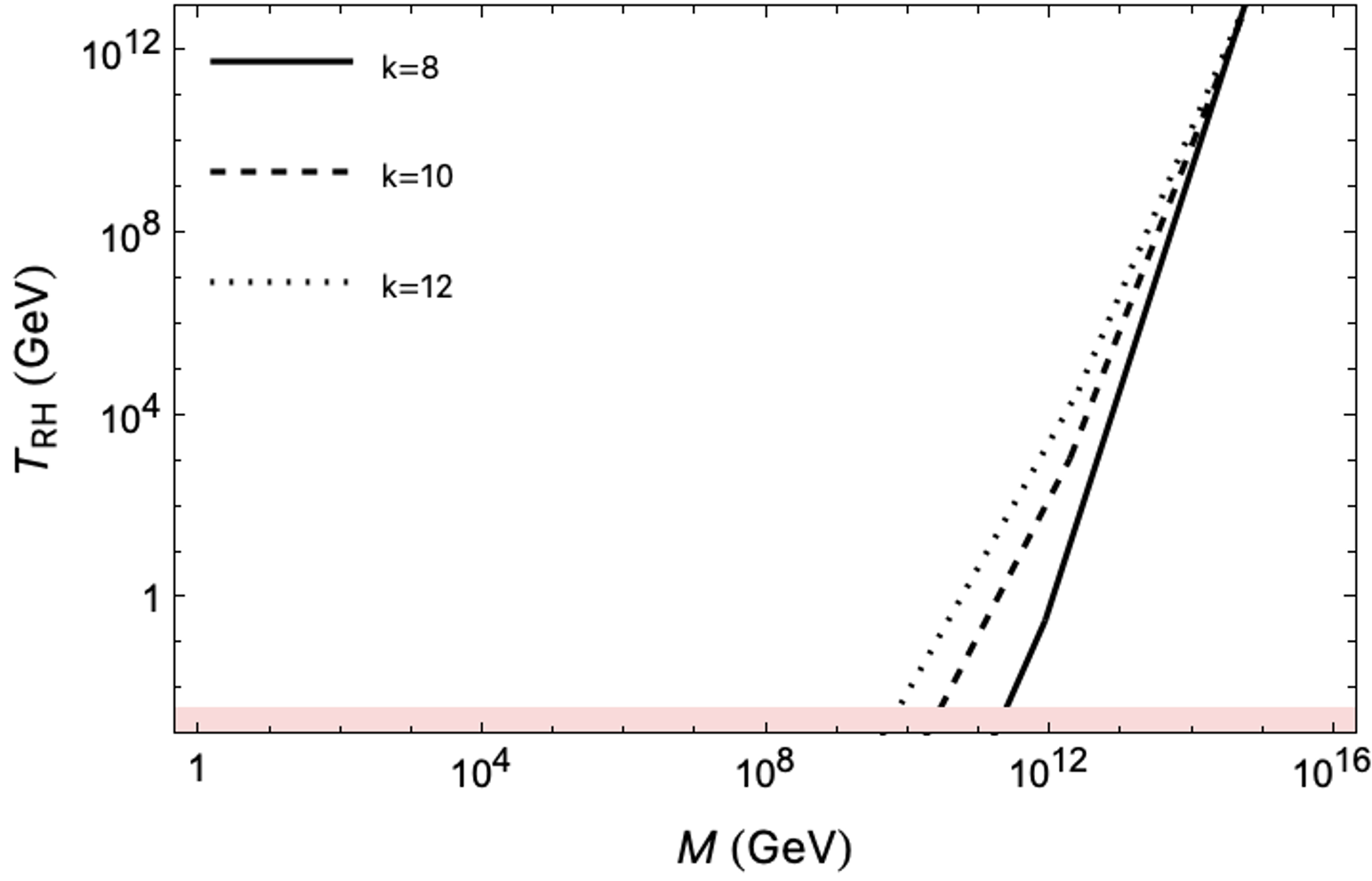}
\caption{\em \small The lower bound on $\Trh$ vs. $M$ for $k=8$, $k=10$, and $k=12$, and $n=2$ with $g=1$, such that $M= \Lambda$. Regions above each line correspond to the interaction reaching equilibrium during the reheating epoch. As a result, relativistic or non-relativistic freeze-out become possible in these regions. Below the lines, the dark component never reaches equilibrium and its abundance is thereby determined by freeze-in. The red shaded region corresponds reheating temperatures below the BBN limit of $T_{\rm RH}=4$ MeV \cite{tr4}.
}
\label{trhCritvsLambda2}
\end{figure}

We show in Fig.~\ref{trhCritvsLambda} the lower bound on $T_{\rm RH}$ such that the interaction will reach equilibrium during reheating as a function of the mediator mass $M$, for $n=2$ and $k=2$, $4$, or $6$. For simplicity, we have taken the dimensionless coupling $g=1$ so that $\Lambda = M$. The regions above the lines correspond to the dark component $\chi$ reaching equilibrium at some point during reheating, thereby making UFO possible. The region where UFO is possible increases with $k$ (for $k < 7$) because for a given reheating temperature, the temperature-dependence of $H(T)$ becomes less steep for large $k$ (see Eq. (\ref{Eq:HT})), which allows $\Gamma(T)$ to exceed $H(T)$ at some time during reheating for larger regions in the $\Trh$ vs $M$ plane.
Below the lines, $\chi$ never enters thermal equilibrium, and its relic density is given by the FI mechanism \cite{HMO}. The lines are computed using Eq.~(\ref{Eq:trhmintmaxkl72}) for $M>\Tmax$ and (\ref{Eq:trhmintmaxkl72g}) for $M < \Tmax$ with $\Tmax$ computed using Eq.~(\ref{tmax}). Note there is a change in slope of $M$ vs $\Trh$ at $M=\Tmax$. For $n=2$ the slope changes from $\Trh \propto M^{\frac{k+2}{2}}$ at small $M$ to $M^{\frac{4k}{5}}$ at larger $M$.  For $M<T_{\rm max}$, the boundary between freeze-in and UFO (treating $T_{\rm RH}$ as a free parameter and keeping $\Lambda$ fixed) will occur at a value of $T_{\rm RH}$ for which the interaction only briefly enters equilibrium near the resonance peak at $T=M$. The results in Figs.~\ref{trhCritvsLambda} and \ref{trhCritvsLambda2} neglect the brief enhancement of $\Gamma(T)$ near the resonance, and therefore provide conservative estimates of the lower bound on $T_{\rm RH}$ that are independent of the width of the mediator.

We show in Fig.~\ref{trhCritvsLambda2} the allowed values of $\Trh$ for $k > 7$. The opposite behavior is seen, and the region where UFO is possible decreases with increasing $k$, since the temperature dependence of $H(T)$ becomes steeper for larger $k$ (see Eq. (\ref{Eq:HTkgeq7})). In this case, the lines are determined by Eq.~(\ref{Eq:trhmintmaxkg72}) $M>\Tmax$ and (\ref{Eq:trhmintmaxkg72g}) for $M < \Tmax$. $\Tmax$ is computed using Eq.~(\ref{tmax2}). 

\subsubsection{Conditions for $\Tfo > \Trh$ and $\Tfo > m_{\chi}$}

Next, we turn to the condition that $\Tfo > \Trh$, namely the requirement that relativistic freeze-out occurs before reheating rather than after reheating. To do this, we can saturate the bound in Eq.~(\ref{tfoConditions}), namely $\Tfo =  \Trh$ so that
\beq
T_{\rm RH} < \left(\frac{3k}{k+2}\frac{\pi^2}{g_{\chi}\zeta(3)}\sqrt{\frac{\alpha}{3}}\frac{\Lambda^{n+2}}{M_P}\right)^{\frac{1}{n+1}}\,,
\label{ultrh}
\eeq
valid for all $k$.

Note that this upper limit on $T_{\rm RH}$ applies in the regime 
where $m_{\chi} < T_{\rm RH}$. In the other regime where $m_{\chi} > T_{\rm RH}$, we must find a separate upper bound on $T_{\rm RH}$ due to the requirement that $T_{\rm FO} > m_{\chi}$ for relativistic freeze-out as in Eq. (\ref{tfoConditions}). When $T_{\rm RH}$ is greater than this value and $m_{\chi} > T_{\rm RH}$, then the DM will undergo non-relativistic freeze-out. Requiring $T_{\rm FO} > m_{\chi}$ leads to the following constraints on $T_{\rm RH}$ necessary for relativistic freeze-out:

For $k<7$ we obtain
\begin{equation}
    T_{\rm RH} < \left(\frac{3k}{k+2}\frac{\pi^2}{g_{\chi}\zeta(3)}\sqrt{\frac{\alpha}{3}}\frac{\Lambda^{n+2}}{M_P}\right)^{\frac{k-1}{2}}m_{\chi}^{\frac{n+3-k-kn}{2}}\,,
    \label{ultrhmgtt}
\end{equation}
whereas for $k>7$
\begin{equation}
    T_{\rm RH} < \left(\frac{3k}{k+2}\frac{\pi^2}{g_{\chi}\zeta(3)}\sqrt{\frac{\alpha}{3}}\frac{\Lambda^{n+2}}{M_P}\right)^{\frac{k+2}{4-k}}m_{\chi}^{\frac{6+n+nk}{4-k}}\,.
\end{equation}

These constraints represent the boundary between relativistic and non-relativistic freeze-out, since when $T_{\rm RH}$ is exceeds these limits, the freeze-out temperature will be below $m_{\chi}$ and therefore freeze-out will be non-relativistic. While these constraints together provide the upper limits on $\Trh$ such that UFO during reheating will occur, recall from the preceding sub-section that the lower limits on $T_{\rm RH}$ are found by setting $T_{\rm FO} = M$, where $M$ is the mass of a heavy mediator (for $M<T_{\rm max}$).

\begin{figure}[!ht]
\centering
\vskip .2in
\includegraphics[width=3.3in]{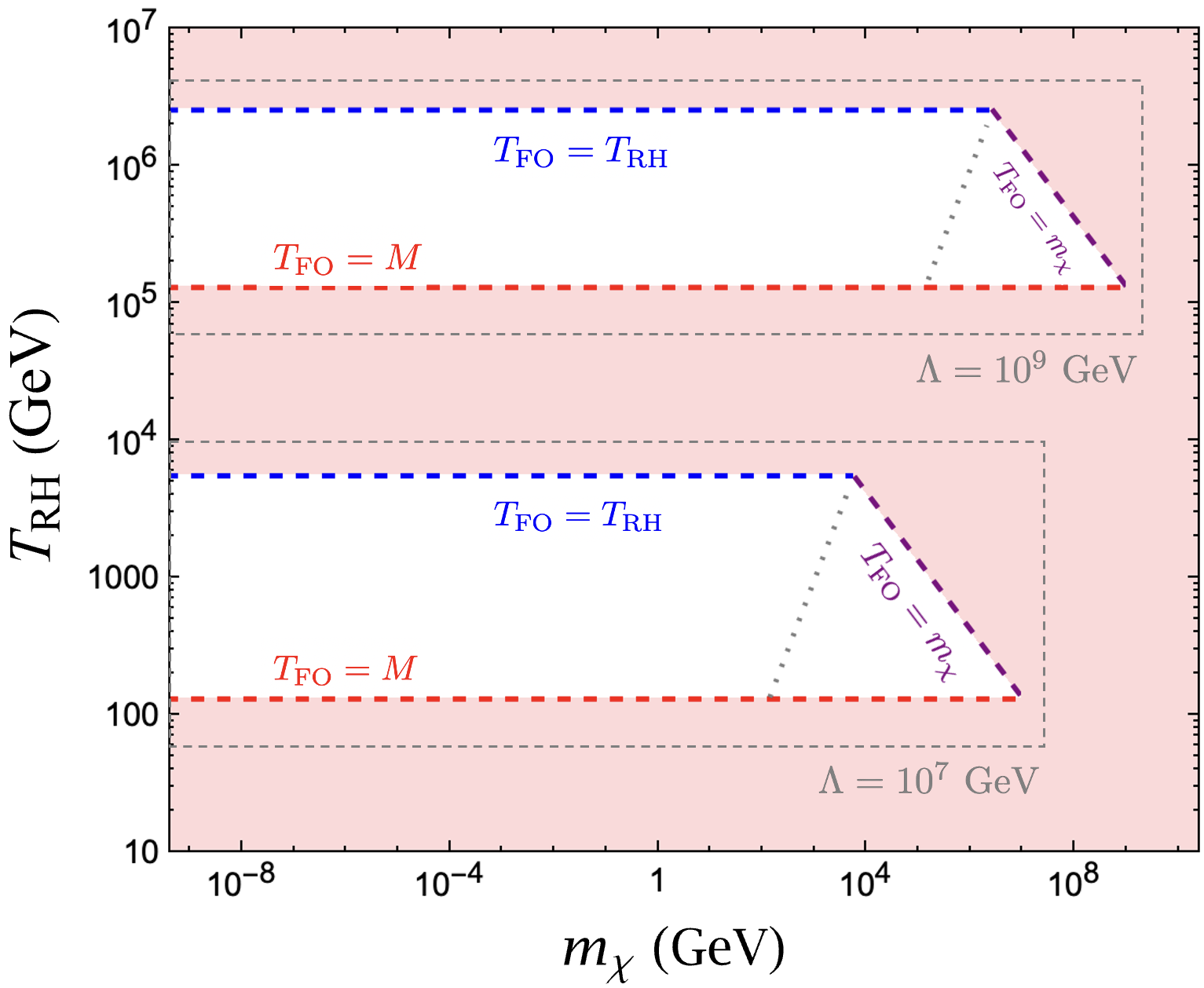}
\caption{The $m_{\chi}$ vs. $\Trh$ plane depicting the constraints on $\Trh$ required for UFO during reheating for $k=2$, $n=2$, and two different values of $\Lambda$. The top portion of the figure depicts the constraints for $\Lambda = 10^9$ GeV, while the bottom portion depicts the constraints for $\Lambda = 10^7$ GeV. In both cases, $\Lambda = \frac{M}{g}$, and we take $g=1$. The white region corresponds to UFO during reheating. The regions above the blue lines for each $\Lambda$ correspond to FO after reheating (during radiation domination). The regions below the red lines for each $\Lambda$ correspond to FI, since the interaction will never reach equilibrium. The region above the purple line corresponds to WIMP-like non-relativistic FO, with $T_{\rm FO} < m_{\chi}$. The gray dotted line is $m_{\chi} = \Trh$.
}
\label{TRHconstraints}
\end{figure}

We have now characterized a set of constraints on $T_{\rm RH}$ for given values of $k$, $n$, and $\Lambda$ that are required for UFO during reheating. In Fig.~\ref{TRHconstraints}, we illustrate these constraints for $k=2$, $n=2$, and two values of $\Lambda = 10^7$ GeV and $10^9$~GeV. The white regions in Fig.~\ref{TRHconstraints} correspond to UFO during reheating, while the red shaded regions correspond to other mechanisms of DM production such as freeze-in, non-relativistic freeze-out during reheating, or freeze-out after reheating. The boundaries of these regions come from Eq.~(\ref{mlttmax22}) (lower limits on $\Trh$) and from setting $\Tfo = \Trh$ in  Eq.~(\ref{Eq:tfok2n2}) or Eq.(\ref{ultrh}) (upper limits on $\Trh$ for  $m_\chi < \Trh$) and (\ref{ultrhmgtt}) (upper limits on $\Trh$ for $m_\chi > \Trh$). The red dashed lines correspond to the approximate UFO/freeze-in boundary, which occurs near $\Tfo=M$. The purple dashed lines correspond to the approximate UFO/non-relativistic FO boundary, near $\Tfo=m_{\chi}$. Lastly, the blue dashed lines correspond to the boundary between UFO during reheating and UFO after reheating, which occurs at $\Tfo=\Trh$.

\section{Evolution of the DM for Ultra-Relativistic Freeze-Out during Reheating}
\label{sec:DMden}

\subsection{Computing the number density $n_\chi$}

\subsubsection{Generalities}

When UFO occurs, 
particles in the Standard Model radiation bath do not produce DM particles at a rate commensurate with the Hubble expansion
($\Gamma(T)< H(T)$). However, inflaton decay continues to steadily increase the co-moving number density of SM particles during reheating. 
This leads the DM number density $n_{\chi}$ to drop relative to the equilibrium number density in the radiation bath, $n_{\rm eq} \propto T^3 \propto a^{-9(k-2)/(2k+4)}$. 
The relevant term on the right-hand side of the Boltzmann equation for relativistic freeze-out is therefore the production term, rather than the annihilation term. That is, DM production may continue to occur after freeze-out via out-of-equilibrium interactions, similar to the freeze-in mechanism.

To understand this better, recall that in a standard FO scenario, prior to freeze-out, equilibrium between the DM and SM is maintained and $n_\chi \propto n_{\rm eq}$ (the constant of proportionality depending on the relevant numbers of degrees of freedom). This comes about from the balance between the source term $\propto n_{\rm eq}^2 \langle \sigma v \rangle$ and the annihilation term $\propto n_{\chi}^2 \langle \sigma v \rangle$ in the Boltzmann equation 
\beq
\frac{dn_\chi}{dt}+3H n_\chi=(n_{\rm eq}^2 -n_\chi^2) \langle \sigma v \rangle\,.
\label{Eq:boltzmann}
\eeq
In the non-relativistic case, as freeze-out is approached the source term becomes negligible compared to the annihilation term as it becomes kinematically unfavorable to produce DM. 
Subsequently, the annihilation term becomes Boltzmann suppressed until it too is negligible compared to the Hubble expansion term, resulting in $n_\chi \propto a^{-3}$ for $a > \afo$.

In the case of UFO,
there is no Boltzmann suppression since freeze-out occurs at temperatures well above $m_{\chi}$. Yet, 
when $n_\chi \langle \sigma v \rangle < H$, $n_{\chi}$ drops relative to the number density of the source particles in the SM bath, and thus the annihilation term in Eq.~(\ref{Eq:boltzmann}) quickly becomes negligible. Nevertheless, the production term, now no longer kinematically suppressed, continues to supply a source of DM as in the FI mechanism via out-of-equilibrium interactions. Thus
we can write the Boltzmann equation after freeze-out (in the relativistic case) as
\beq
\frac{dn_\chi}{dt}+3H n_\chi=n_{\rm eq}^2 \langle \sigma v \rangle\,,
\label{Eq:boltzmannfo}
\eeq
for $T<\Tfo$ with an initial abundance of DM given by $n_\chi(\Tfo)$ when the out-of-equilibrium dynamics begin. 
Unlike the case of non-relativistic freeze-out,
$n_\chi$ is not simply proportional to $a^{-3}$ during the phase of reheating for $a > \afo$. 
As in the FI mechanism, 
dark matter may be produced after $\afo$, 
even if $n_{\rm eq}\langle \sigma v \rangle\ll H$. This is because 
 the {\it number} of targets in the production rate, $n^2_{\rm eq}\langle \sigma v \rangle \propto a^3n_{\rm eq}$ can increase 
with time, because SM particles are still steadily produced by inflaton decay.

To determine the DM number density it is convenient to solve the Boltzmann equation in terms of the comoving number density $Y_\chi=n_\chi a^3$. 
 Eq.~(\ref{Eq:boltzmannfo}) can be rewritten as
\begin{equation}
\frac{dY_{\chi}}{da} = \frac{a^2 \langle \sigma v \rangle n_{\rm eq}^2}{H(a)}\,.
\label{Boltzmann1}
\end{equation}
It will be convenient to compare the derived abundance $Y_\chi$, with
\beq
Y_{\rm eq} \equiv \frac{g_\chi \zeta(3)}{\pi^2} T^3 a^3 \, .
\eeq
Note that only for $T \gg m_\chi$, is $Y_{\rm eq} \simeq n_{\rm eq} a^3$.
In a standard freeze-out scenario (with freeze-out after reheating), the universe is in 
an isentropic radiation era where $T\propto a^{-1}$, or 
constant $a^3n_{\rm eq}$.  For $\langle \sigma v \rangle \propto T^n$ and $H \propto a^{-2}$, Eq.~(\ref{Boltzmann1}) gives  $\frac{dY_\chi}{da}\propto a^{-(n+2)}$. In this case, the amount of dark 
matter produced after $\afo$ is negligible for $n>-1$, as in the case of neutrino decoupling. Of course, for $a>\arh$, the comoving number 
$Y_\chi$
stays constant because $n_{\rm eq}\propto a^{-3}$ after the end of the reheating process.
In contrast with the more standard scenario, the relation between $T$ and $a$ differs during reheating. From Eq.~(\ref{Eq:rhoR}), we see that $T \propto a^{-3(k-1)/2(k+2)}$
or $n_{\rm eq}\propto a^{-\frac{9}{8}}$ for $k=2$. In this case, $\frac{dY_\chi}{da}\propto a^{-\frac{3}{8}n+\frac{5}{4}}$ and the late production of DM is only negligible for $n>6$.

We are now in a position to compute the final relic density of DM in the UFO scenario. We distinguish the two cases with $m_\chi < \Trh$ and $m_\chi > \Trh$.

\subsubsection{$m_\chi < \Trh$}

Starting with $k<7$, we can 
use Eq.~(\ref{Eq:rhoR}) to obtain $T(a)$ and Eq.~(\ref{Eq:rhophi}) with $H = \sqrt{\rho_\phi}/3M_P$, 
so that 
\begin{equation}
\frac{dY_{\chi}}{da} \propto a^{\frac{3n+26-8k-3kn}{2k+4}}\,.
\label{Boltzmann2}
\end{equation}
This is easily integrated 
from  $\afo$ to some desired scale $a$ after freeze-out
\begin{align}
Y_{\chi}(a) &= Y_{\rm FO} + \frac{g_{\chi}^2 \zeta(3)^2}{ \pi^4}\sqrt{\frac{3}{\alpha}}\frac{T_{\rm RH}^{n+4}M_{P}}{\Lambda^{n+2}} a_{\rm RH}^{\frac{(3k-3)(n+6)-6k}{2k+4}} \nonumber \\ & \times \left(\frac{2k+4}{3n-3nk-6k+30}\right)\left[a^{\frac{(3-3k)(n+6)+12k+12}{2k+4}}- \nonumber \right. \\ & \left. a_{\rm FO}^{\frac{(3-3k)(n+6)+12k+12}{2k+4}}\right].
\label{generalYa}
\end{align}
For $m_{\chi} < \Trh$, the IR scale we should integrate to is $a = \arh$, since, depending on the value of $n$, DM production may continue to occur after freeze-out from $a_{\rm FO}$ through $a_{\rm RH}$ in this case. With this limit of integration, can rewrite the Eq.~(\ref{generalYa}) as
\bea
&&Y_\chi(\arh) = \Yfo + \frac{ g_\chi^2\zeta(3)^2}{ \pi^4}  \left( \frac{2k+4}{3n-3nk-6k+30} \right)   \times
\nonumber \\ 
&& \sqrt{\frac{3}{\alpha}}\frac{T_{\rm RH}^{n+4}M_{P}}{\Lambda^{n+2}} \left[
\arh^3
- \afo^3\left(\frac{\afo}{\arh}\right)^{\frac{3n-3nk+18-12k}{2k+4}}\right]\,.
\label{Eq:yrh}
\eea
The value of $\Yfo$ can be determined from Eqs.~(\ref{ntfo}) and (\ref{Eq:tfokl7}) (for $k<7)$,
\begin{align}
\Yfo = Y_{\chi}(a_{\rm fo}) = & n_{\chi}(a_{\rm fo})a_{\rm fo}^3  \nonumber \\
 = & \frac{g_{\chi} \zeta(3)}{\pi^2}T_{\rm fo}^3 a_{\rm RH}^3 \left(\frac{T_{\rm RH}}{T_{\rm fo}}\right)^{\frac{2k+4}{k-1}}  \nonumber \\
 = & \frac{g_{\chi} \zeta(3)}{\pi^2}\left[\frac{3k}{k+2} \sqrt{\frac{\alpha}{3}}\frac{\pi^2 \Lambda^{n+2}}{g_{\chi} \zeta(3) M_P }\right]^{\frac{k-7}{nk -n+k-3}} \nonumber \\ 
 \times & T_{\rm RH}^{\frac{2(kn+2n+k-1)}{nk -n+k-3}} \arh^3\,.
 \label{Eq:yfo}
\end{align}

For each interaction type (i.e. for a given value of $n$), the out-of-equilibrium production will occur predominantly in either the IR or UV regime. That is, the out-of-equilibrium contribution to the comoving number density $Y_{\chi}$ will be determined predominantly by either the first or second term in the large brackets in Eq.~(\ref{Eq:yrh}), respectively.  There is a critical value of $n$ that separates the two, given by $n^{*} = \frac{10-2k}{k-1}$ \cite{gkmo1}. For $n > n^{*}$, the interaction will be UV-dominated. 
In this case, the comoving
number $Y_\chi$ does not depend on the scale factor $a$ but only on the 
conditions at $\afo$.
In contrast, for $n < \frac{10-2k}{k-1}$, the out-of-equilibrium production is IR dominated. The comoving number increases with the scale factor, and its value at $\arh$
no longer depends on the physics at $\afo$. In the latter case, it matters, however,  whether $m_{\chi}$ is greater than or less than $T_{\rm RH}$. 

It is worth noting that this result is very similar to what we 
found for the FI mechanism \cite{HMO}. In fact, the 
solution is identical apart from the limits of integration and 
the initial condition, $\Yfo$. For FI, we integrate from $a_{\rm end}$ to some 
desired larger value of $a$ (often $a_{\rm RH}$), but the initial 
condition is typically taken to be $Y_{\chi}(a_{\rm end}) = 0$, 
since the interaction never reaches equilibrium for standard FI. In 
contrast, for UFO, we begin from the 
equilibrium condition at freeze-out, $Y_\chi(\afo) $, and subsequently compute the evolution of the 
number density. Despite the similarity of the UFO and FI solutions to the Boltzmann equation, we will see that the UFO parameter space is quite different from that of standard FI.

\begin{figure}[!ht]
\centering
\vskip .2in
\includegraphics[width=3.3in]{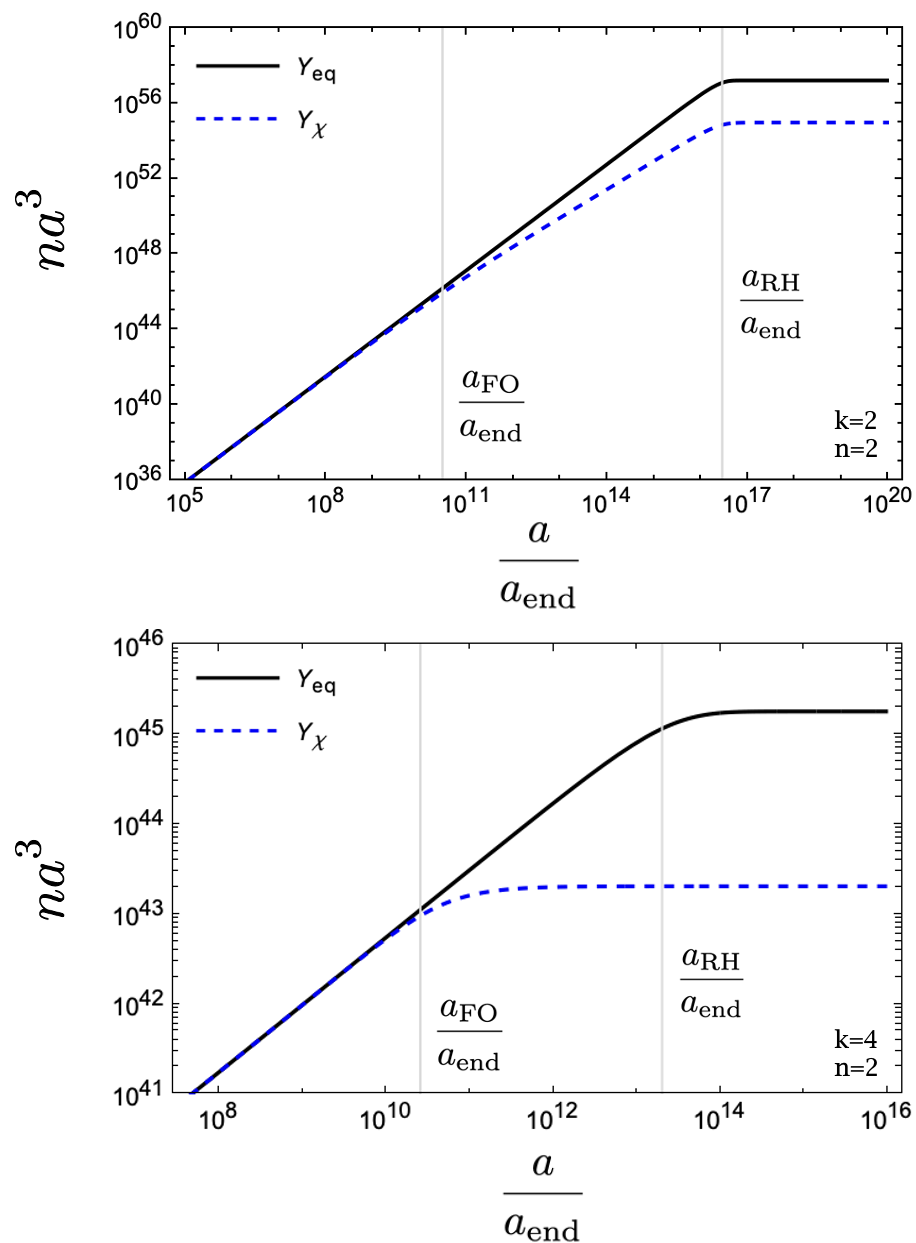}
\caption{\em \small Evolution of the co-moving DM number density ($Y_{\chi} = n_{\chi}a^{3}$) for relativistic freeze-out during reheating as a function of the scale factor ($a$) for IR (top panel) vs. UV (bottom panel) production and $m_{\chi} < T_{\rm RH}$. The parameter choices for each are as follows.\\ \hspace{10 mm} \textbf{Top:} $k=2$, $n=2$, $\Lambda = 10^{7} \text{ GeV}$, $T_{\rm RH} = 1000 \text{ GeV}$.\\ \textbf{Bottom:} $k=4$, $n=2$, $\Lambda = 10^{7} \text{ GeV}$, $T_{\rm RH} = 100 \text{ GeV}$.}
\label{comovingNumDens}
\end{figure}

The evolution of the comoving number density is shown in Fig.~\ref{comovingNumDens} for $n=2$ and $\Lambda = 10^7$~GeV. 
We have taken $k=2$ ($k=4$)  and $\Trh = 1000$ GeV ($\Trh=100$ GeV) in the upper (lower) panel. For $k=2$ and $n=2$, post freeze-out production of DM is IR dominated.
Its evolution is shown by the blue dashed line in Fig.~\ref{comovingNumDens}. This is compared with the evolution of $Y_{\rm eq} \propto a^{\frac{15}{8}}$ until $a = \arh$. 
For $a > \arh$, $n_{\rm eq} \sim a^{-3}$ and $Y_{\rm eq}$ is constant. The value of $\arh/\aend$ is marked by the right vertical line. At $a = \afo$ (marked by the left vertical line), $Y_{\chi} < Y_{\rm eq}$ as $Y_\chi$ scales as $a^{3/2}$ in this case. For $m_\chi < \Trh$, $Y_\chi$ is also constant for $a > \arh$.

From Eq.~(\ref{Eq:TFO}), the freeze-out temperature for the case with $k=2, n=2$, $\Trh=1000$~GeV, $\Lambda = 10^7$~GeV, and $g_\chi = 1$ is $\Tfo\simeq 1.7 \times 10^5$ GeV and we can neglect\footnote{As we 
explained earlier, in the IR case, this comes from the fact 
that $n_{\rm FO}$ redshifts faster than $(\Trh/\Tfo)^3$.} $\Yfo$ at $\arh$, and write, with $Y_{\chi}(\arh)=\Yrh$,
\bea
&&\frac{\Yrh}{n_\chi^{\rm eq}(\arh) \arh^3}=\frac{\Yrh}{Y_{\rm eq}(\arh)}
\simeq\frac{2}{\sqrt{3 \alpha}}
\frac{n_\chi^{\rm eq}(\arh)M_P}{\Lambda^4}
\nonumber
\\
&&\simeq 5.8\times 10^{-3}\left(\frac{10^7}{\Lambda}\right)^4\left(\frac{\Trh}{1000~\rm GeV}\right)^3 \,.
\label{Eq:ratiok2n2}
\eea
 This corresponds to the ratio of the number of dark matter particles over the relativistic species in the bath (up to some degrees of freedom
squared),
which is constant after reheating.
This is what we observe in the upper panel of Fig.~\ref{comovingNumDens}.

Next, we turn to the lower panel of Fig.~\ref{comovingNumDens}, where we depict the evolution of $Y_{\chi}(a)$ for $k=4$ and $n=2$, with all other parameters fixed at their previous values. Note that this corresponds to $n > n^{*}$, such that the out-of-equilibrium production is UV dominated, and therefore the second term in large brackets in Eq.~(\ref{Eq:yrh}) dominates. The production of DM from the thermal bath in this case is much less efficient than IR domination, due to the dilution of the source, the inflaton, which behaves as radiation ($\rho_\phi\propto a^{-4}$). Thus, freeze-out during reheating occurs in a ``radiation dominated" era for $k=4$, as in the case of classical relativistic freeze-out. However, freeze-out during reheating with $k=4$ still yields drastically different results from the classical case, since the DM undergoes an extra period of dilution from $a_{\rm FO}$ to $a_{\rm RH}$, which allows for a much broader set of $\Lambda$, $m_{\chi}$, and $T_{\rm RH}$ values to be compatible with $\Omega_{\chi} h^2 = 0.12$, as we will see. 

It is useful in the UV case to re-write Eq.~(\ref{Eq:yrh}) in the following way, which will lead to a simplification:
\beq
Y_{\chi}(a_{\rm RH})=Y_{\rm FO}+Y_{\rm IR} + Y_{\rm UV},
\label{Eq:YRHsimplified}
\eeq
\noi where $Y_{\rm IR}$ and $Y_{\rm UV}$ are the second and third terms in Eq.~(\ref{Eq:yrh}) respectively, representing the post-freeze-out production. We can then simplify this expression further by demonstrating that $Y_{\rm UV}$ is simply a rational multiple of $Y_{\rm FO}$, as follows: 
\begin{align}
Y_{\rm UV} & = \frac{ g_\chi^2\zeta(3)^2}{ \pi^4}  \left( \frac{2k+4}{3nk-3n+6k-30} \right) \sqrt{\frac{3}{\alpha}}\frac{T_{\rm RH}^{n+4}M_{P}}{\Lambda^{n+2}} \nonumber \\ &\times \left[\afo^3\left(\frac{\afo}{\arh}\right)^{\frac{3n-3nk+18-12k}{2k+4}}\right] \nonumber \\
&=\frac{g_{\chi} \zeta(3)}{\pi^2}T_{\rm FO}^3 a_{\rm FO}^3 \left( \frac{2k+4}{3nk-3n+6k-30} \right)\left(\frac{3k}{k+2}\right) \nonumber \\ &= \left( \frac{2k}{nk-n+2k-10} \right) Y_{\rm FO}.
\label{Eq:YUVderivation}
\end{align}
Substituting this result into Eq.~(\ref{Eq:YRHsimplified}) then gives
\beq
Y_{\chi}(a_{\rm RH})=\left[ 1\!+\left( \frac{2k}{nk-n+2k-10} \right)\right]Y_{\rm FO}+Y_{\rm IR},
\label{Eq:YRHforUV}
\eeq
which for $k=4$ and $n=2$ becomes
\beq
Y_{\chi}(a_{\rm RH})=3 Y_{\rm FO}+Y_{\rm IR} \approx 3 Y_{\rm FO}.
\eeq
Therefore, we find 
\bea
&&
\frac{\Yrh}{Y_{\rm eq}(\arh)}=\frac{3\Yfo}{Y_{\rm eq}(\arh)}=3\frac{\Tfo^3}{\Trh^3}\frac{\afo^3}{\arh^3}=3\frac{\Trh}{\Tfo}
\nonumber
\\
&&
= 3 \left(\frac{g_\chi\zeta(3)}{2 \pi^2}\sqrt{\frac{3}{\alpha}}\right)^\frac37\frac{\Trh^{\frac{9}{7}} M_P^\frac37}{\Lambda^\frac{12}{7}}
\label{Eq:ratio1}
\\
&&
\simeq 1.5 \times 10^{-2} \left(\frac{\Trh}{100~\rm GeV}\right)^{9/7}
\left(\frac{10^7~\rm GeV}{\Lambda}\right)^{\frac{12}{7}}
\nonumber
\eea
 where we used $\Tfo$ from Eq.~(\ref{Eq:tfokl7}), $a\propto T^{-\frac43}$ from Eq.~(\ref{Eq:rhoR}), and $g_\chi = 1$. This can be seen in the lower panel of Fig.~\ref{comovingNumDens}, obtained from numerically solving the set of Boltzmann equations.

Further increasing the value of $k$ causes $\rho_\phi(a)$ to redshift faster (see Eq.~(\ref{Eq:rhophi})), rendering $\arh$ smaller for fixed $\Trh$.  This
reduces the relative values $\afo$ and $\arh$. The comoving number for $\chi$ is then nearer to the comoving number at equilibrium compared to the case $k=4$.

Increasing the value of $n$ also modifies the evolution 
of the comoving number density of $\chi$. The production is less efficient over time for larger $n$, and we expect a lower comoving number density at $\arh$.
For $k=2$, $n^* = 6$, so 
for $n=4$ post freeze-out production is still IR dominated. However, now 
$Y_\chi(a)\propto a^{\frac34}$.
The slope is less steep than for $n=2$ because the cross section $\langle \sigma v \rangle \propto T^n$ is less efficient
for lower $T$. In other words, for $n=4$, dark matter is less efficiently produced over time from scattering within the thermal bath than for $n=2$. 

\begin{figure}[!ht]
\centering
\vskip .2in
\includegraphics[width=3.3in]{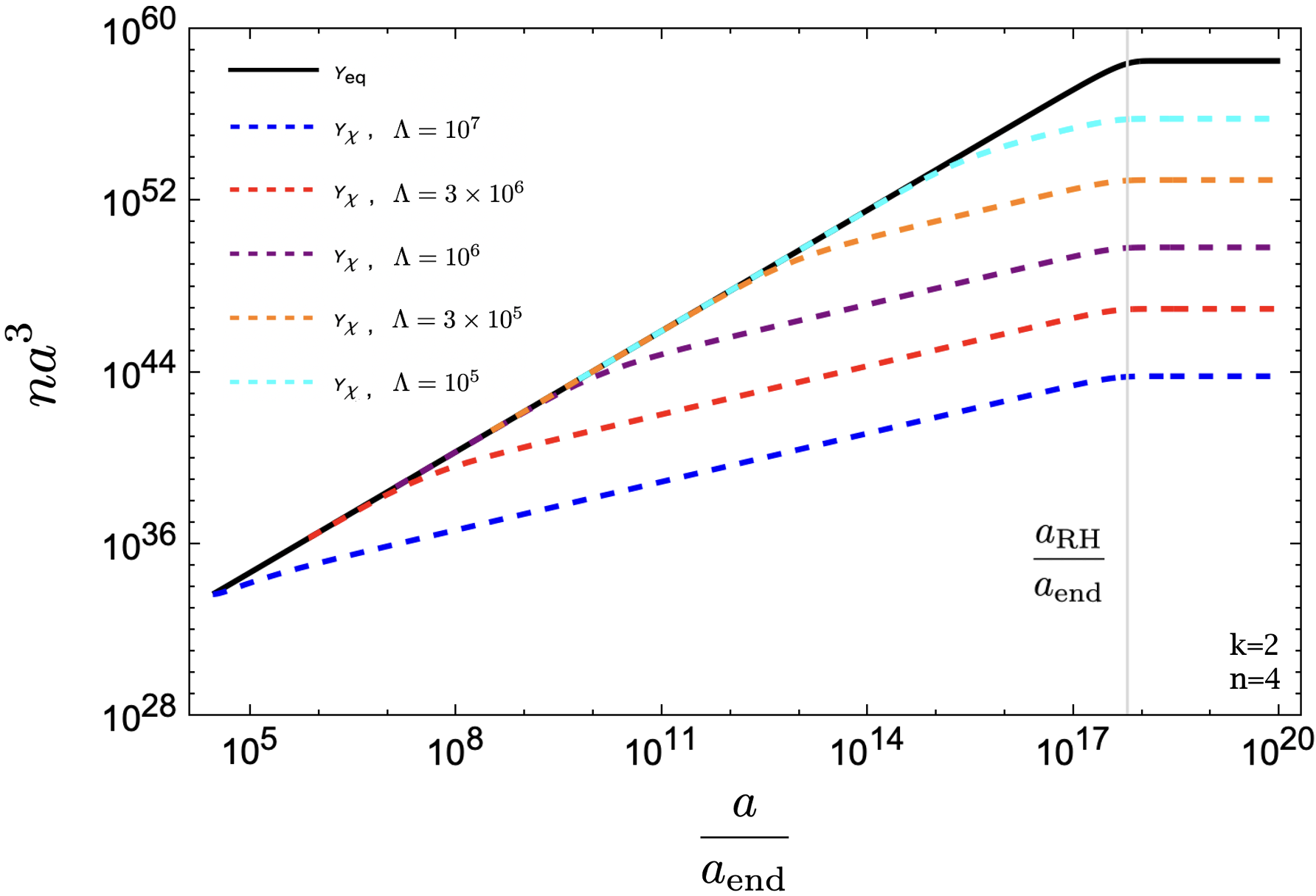}
\caption{\em \small $\Lambda$-dependence of $Y_{\chi}$ for relativistic freeze-out during reheating mediated by an IR dominated interaction. The parameter values are: $k=2$, $n=4$, $T_{\rm RH} = 100 \text{ GeV}$ and $m_{\chi} = 100$ MeV. 
}
\label{YLambda1}
\end{figure}

We show in Fig.~\ref{YLambda1} the evolution of $Y_\chi$ as 
function of $a$ for $k=2$, $n=4$, $m_\chi = 100$~MeV, and $\Trh=100$ GeV for 
different values of $\Lambda$. The slopes of the dashed lines are all 
$Y_\chi\propto a^\frac34$ as expected.  We can further quantify the dependence of $Y_\chi$ on $\Lambda$. From Eq.~(\ref{Eq:yrh}), keeping only the first (IR) term, we obtain
\bea
&&
\frac{\Yrh}{Y_{\rm eq}(\arh)}\simeq\frac{4 g_\chi\zeta(3)}{\pi^2\sqrt{3 \alpha}}\frac{\Trh^5M_P}{\Lambda^6}
\\
&&
\simeq 1.2\times 10^{-15}\left(\frac{\Trh}{100~\rm GeV}\right)^5\left(\frac{10^7~\rm GeV}{\Lambda}\right)^6\,,
\nonumber
\eea
which is effectively what we  observe in Fig.~\ref{YLambda1}.
Comparing to the case with $n=2$, Eq.~(\ref{Eq:ratiok2n2}), 
we notice, a large depletion of the relic abundance, by about 9 orders of magnitude!

\begin{figure}[!ht]
\centering
\vskip .2in
\includegraphics[width=3.3in]{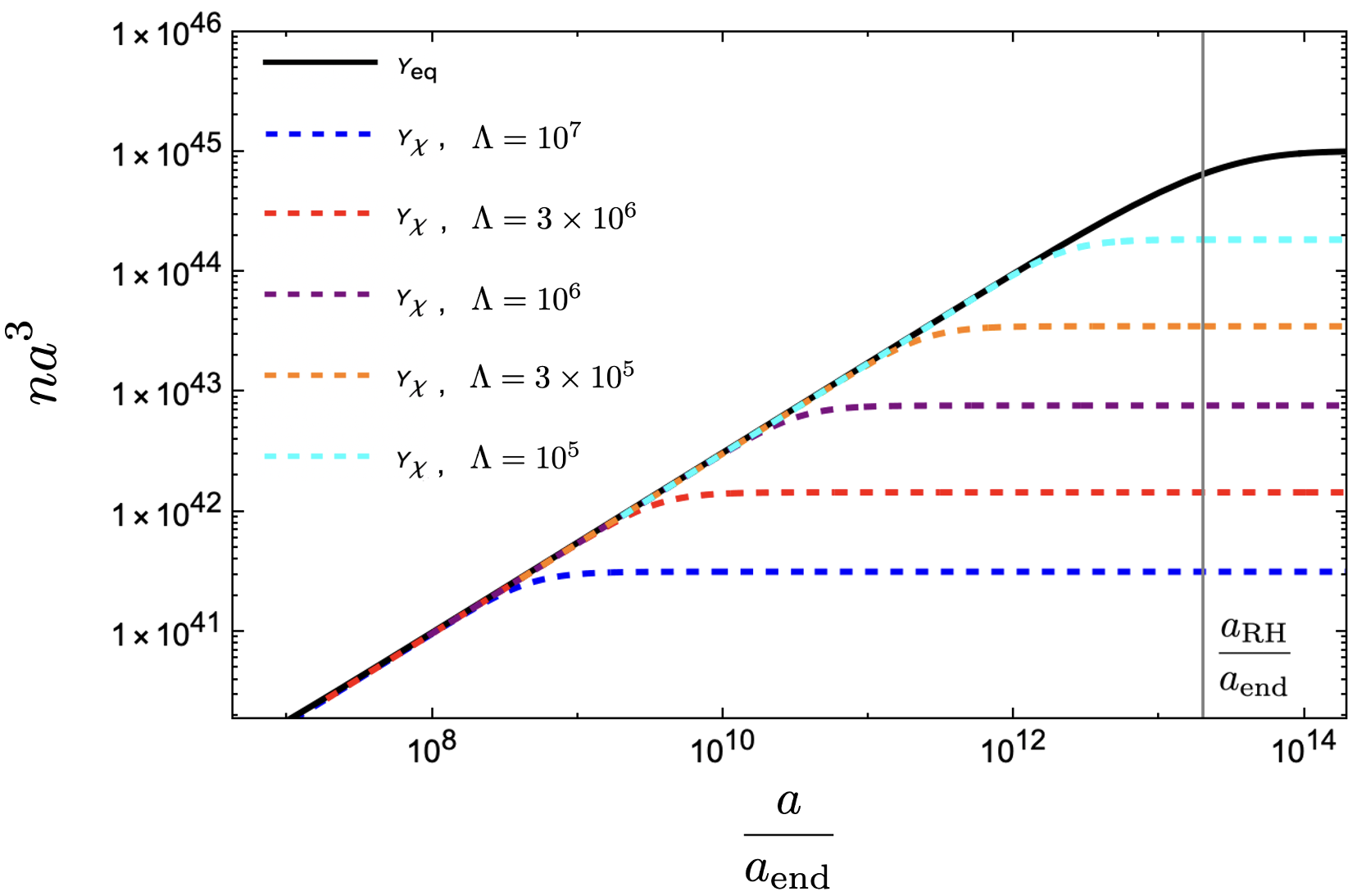}
\caption{\em \small $\Lambda$-dependence of $Y_{\chi}$ for relativistic freeze-out during reheating mediated by a UV dominated interaction. The parameter values are: $k=4$, $n=4$, $T_{\rm RH} = 100 \text{ GeV}$, and $m_{\chi} = 100 \text{ MeV}$. 
}
\label{YLambda2}
\end{figure}

In Fig.~\ref{YLambda2}, we plot $Y_\chi(a)$ for $k=4$ and $n=4$, with $\Trh=100$ GeV
and different values of $\Lambda$. 
Production in this case is UV dominated, and from Eq.~(\ref{Eq:yrh}), 
the dominant term  is given simply by the term proportional to $Y_{\rm FO}$ in Eq.~(\ref{Eq:YRHforUV}), and we obtain
\begin{align}
    \frac{\Yrh}{Y_{\rm eq}(\arh)}
&=\frac{9}{5}\frac{T_{\rm RH}}{T_{\rm FO}}=\frac{9}{5}\frac{\Trh^{\frac{15}{13}} M_P^\frac{3}{13}}{\Lambda^{\frac{18}{13}}}
\left(\sqrt{\frac{3}{\alpha}}\frac{g_\chi \zeta(3)}{2 \pi^2}\right)^\frac{3}{13} \nonumber \\ & \simeq 5.1 \times 10^{-4} \left(\frac{T_{\rm RH}}{100 \text{ GeV}}\right)^{15/13}\left(\frac{10^7 \text{ GeV}}{\Lambda}\right)^{18/13}
\end{align} 
for $k=4$, $n=4$, and $g_\chi = 1$ which is effectively what we observe in Fig.~\ref{YLambda2}. As expected, due to the lack of efficiency in the production process, the comoving number is reduced compared to the case $n=2$ obtained in Eq.~(\ref{Eq:ratio1}).

\subsubsection{$m_\chi> \Trh$}

If the dark matter mass is higher than the reheating temperature, 
the IR freeze-in production stops at a temperature greater than
$\Trh$. Indeed, when $T<m_\chi$, the exponentially suppressed Boltzmann factor significantly reduces the number of targets in the
$n_{\rm eq}\times n_{\rm eq}\langle \sigma v \rangle$ production term. In this case, 
we should integrate Eq.~(\ref{Boltzmann1}) from $\aend$ to $\am$,
where $\am$ is defined by $T=m_\chi$. 
Note that this will in general lead to an over-estimate in our analytic expressions unless the Boltzmann suppression in the production rate is included. It is included in our numerical calculations. 
Clearly,  the relative magnitude 
between $\Trh$ and $m_\chi$ does not play any role in the case of UV production, when $n>n^*$ and $\Yfo$ is determined at $\afo$.

\begin{figure}[!ht]
\centering
\vskip .2in
\includegraphics[width=3.3in]{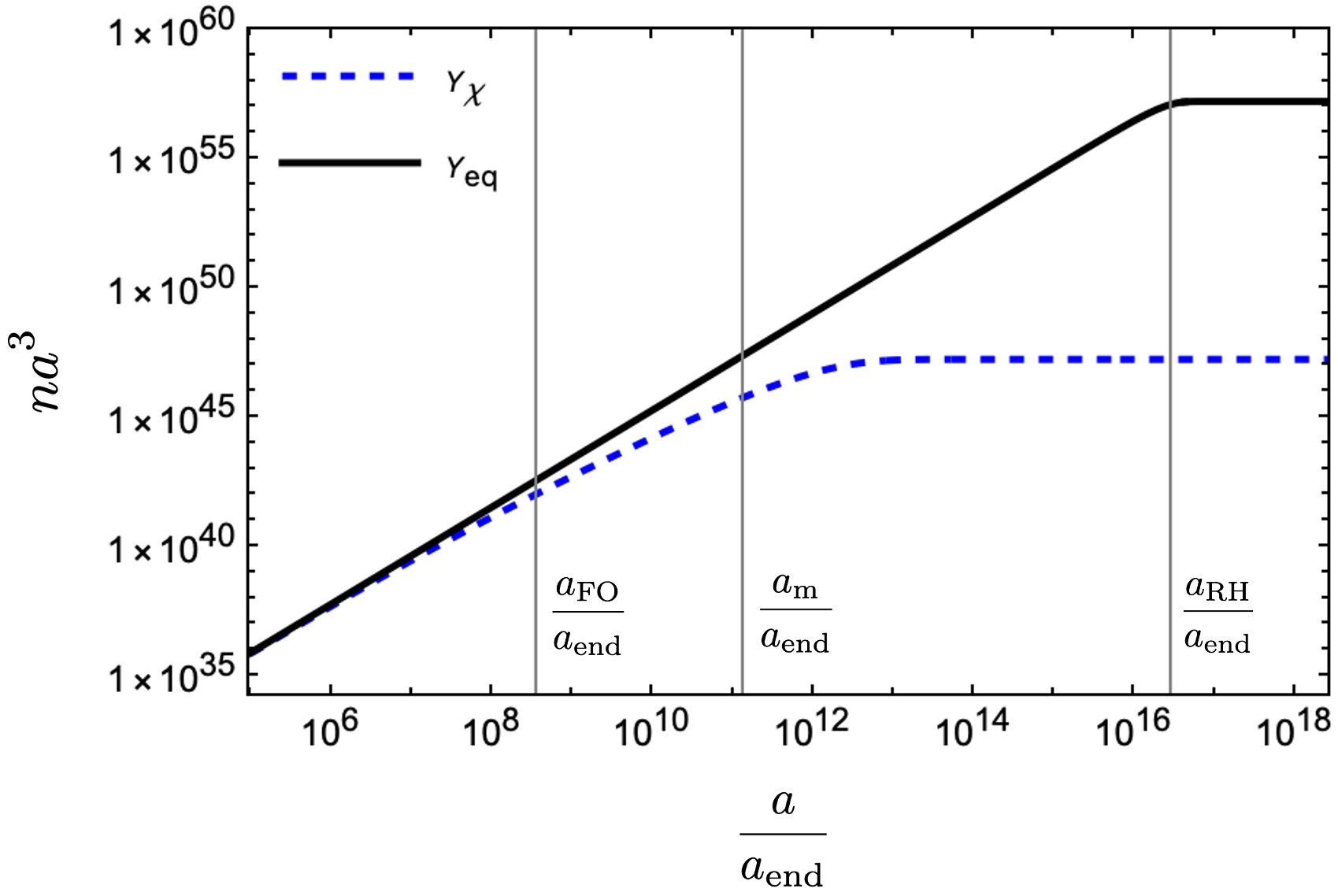}
\caption{\em \small Evolution of the co-moving DM number density ($Y_{\chi} = n_{\chi}a^{3}$) for UFO during reheating as a function of the scale factor ($a$) for IR production and $m_{\chi} > T_{\rm RH}$. The parameter choices are as follows: $k=2$, $n=2$, $\Lambda = 2 \times 10^{7} \text{ GeV}$, $T_{\rm RH} = 1000 \text{ GeV}$, and $m_{\chi} = 10^{5} \text{ GeV}$.}
\label{comovingNumDens2}
\end{figure}

The integration of Eq.~(\ref{Boltzmann1}) between $\afo$ and $\am$ keeping only the IR part with $k<7$ gives,
\bea
&&\frac{\Yrh}{Y_{\rm eq}{(\arh)}}
\simeq \frac{Y_\chi(\am)}{Y_{\rm eq}(\arh)} \simeq
\frac{g_\chi\zeta(3)}{\pi^2}\sqrt{\frac{3}{\alpha}}
\frac{m_\chi^{n+4}M_P}{\Lambda^{n+2}\Trh^3}
\nonumber
\\
&& \times \frac{2k+4}{3n-3nk-6k+30}\left(\frac{\am}{\arh}\right)^{\frac{3(k+3)}{k+2}}
\label{Eq:ratiomchi}
\\
&&\simeq\frac{g_\chi\zeta(3)}{\pi^2}\sqrt{\frac{3}{\alpha}}
\frac{(2k+4)~m_\chi^\frac{nk-n+2k-10}{k-1}}{3n-3nk-6k+30}
\frac{\Trh^{\frac{9-k}{k-1}}M_P}{\Lambda^{n+2}}\,,
\nonumber
\eea
where we used, from Eq.~(\ref{Eq:rhoR})
\beq
\frac{\am}{\arh}\simeq\left(\frac{\Trh}{m_\chi}\right)^\frac{2k+4}{3k-3}
\,,
\eeq
and we neglected $\Yfo\ll Y_\chi(\am)=\Yrh$. For $k=2$ and $n=2$,
Eq.~(\ref{Eq:ratiomchi}) gives
\bea
&&
\frac{\Yrh}{Y_{\rm eq}{(\arh)}} \simeq \frac{2 g_\chi\zeta(3)}{\pi^2\sqrt{3 \alpha}}\frac{\Trh^7 M_P}{\Lambda^4 m_\chi^4}
\label{Eq:ratiomchik2n2}
\\
&&
\simeq
3.6\times 10^{-12}\left(\frac{\Trh}{1000~\rm GeV}\right)^7\left(\frac{2\times 10^7~\rm GeV}{\Lambda}\right)^4 \nonumber \\ 
&& \times \left(\frac{10^5~\rm GeV}{m_\chi}\right)^4\,.
\nonumber
\eea
This is somewhat smaller (by a factor of $\sim$ 30) than what is seen in Fig.~\ref{comovingNumDens2}, which can be compared with the upper panel of Fig.~\ref{comovingNumDens}.
Recall that in Eq.~(\ref{Eq:ratiomchik2n2}), we have integrated the Boltzmann equation down to $T=m_\chi$
always assuming the same production rate for a relativistic particle. Near $\am$, this approximation breaks down, and in an IR dominated case, this is manifest as an over-estimate of $n_\chi(\arh)$, since we might have more accurately cut off the integration at $T > m_\chi$. However, because we are plotting $Y$,
there is an additional enhancement by $a^3$ which overcompensates and the net result of the two approximations accounts for the difference between the numerical and analytical results when comparing with Fig.~\ref{comovingNumDens2}.
Recall also that $Y_{\rm eq}$ is defined in terms of a massless particle, but this is used only as a baseline for comparison and does not affect the difference between the numerical and analytical results. We stress again that numerical curve in Fig.~\ref{comovingNumDens2} does take into account the fact that $\chi$ becomes non-relativistic near $\arh$.

We clearly see the departure from the equilibrium at $\afo$,
where $Y_\chi(a)$ tends to follow an evolution $\propto a^\frac32$
before being frozen after $\am$.
As a consequence, we observe
a large depletion of the relic abundance of $\chi$, comparing
Eqs.~(\ref{Eq:ratiomchik2n2}) and (\ref{Eq:ratiok2n2}).
This comes from the fact that the freeze-out happens much earlier 
than $\arh$, limiting substantially the production process.

All of the above can be repeated for $k>7$
where the behavior of $\rho_{\rm R}$ now scales simply as $a^{-4}$ as in Eq.~(\ref{Eq:rhoR}). We leave this as an exercise to the motivated reader.

\subsection{Determining the Relic Abundance}

Having determined the expressions of $\Yrh$, in Eq.~(\ref{Eq:yrh}), it is relatively easy to determine the present-day
relic abundance of $\chi$.

The fraction of the present critical density can be obtained directly from from $\Yrh$ \cite{mybook}
\begin{align}
\frac{\Omega_\chi h^2}{0.12} & = 4.9 \times 10^7~
\frac{\Yrh}{\arh^3\Trh^3} \frac{m_\chi}{{\rm GeV}} \nonumber \\ 
& = 6.0 \times 10^6 \frac{\Yrh}{Y_{\rm eq}{(\arh)}} \frac{m_\chi}{{\rm GeV}} \,, 
\label{Omega}
\end{align}
for $g_\chi = 1$.
Plugging Eq.~(\ref{Eq:yrh}) or any of the ratios, ${\Yrh}/{Y_{\rm eq}{(\arh)}}$, into  (\ref{Omega}), we obtain the following relic abundances for several cases with IR or UV domination and specific values of $k$.

\begin{figure}[!ht]
\centering
\vskip .2in
\includegraphics[width=3.3in]{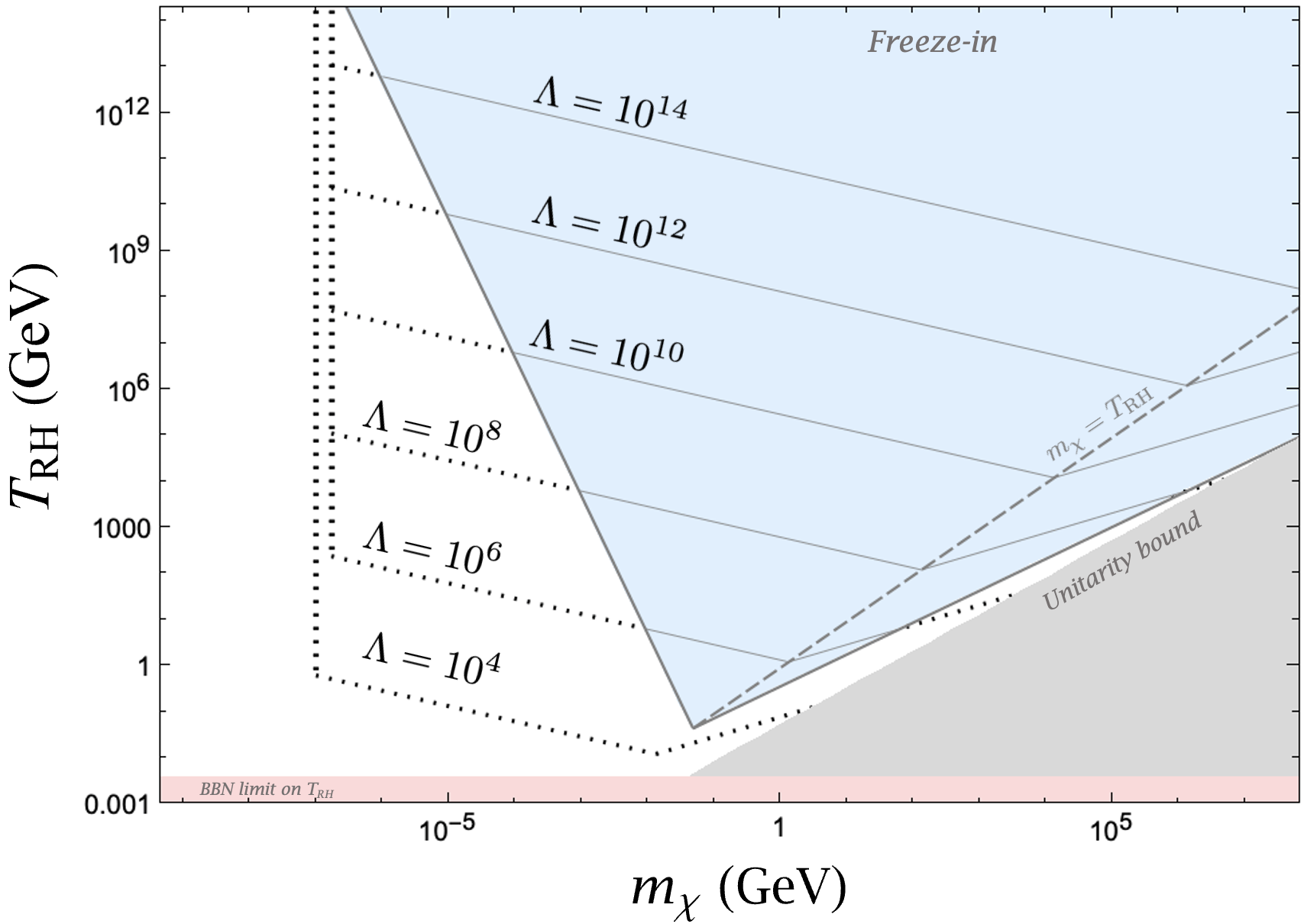}
\caption{\em \small The $m_{\chi}$ vs. $T_{\rm RH}$ plane consistent with $\Omega_{\chi} h^2=0.12$ for $k=2$ and $n=2$ with $g=1$, so that $\Lambda=M$. The solid dark gray lines corresponding to $\Tfo = M$ determine the boundary between FI and FO. The white region corresponds to UFO for each $\Lambda$ listed. The thin gray lines in the blue region correspond to the correct relic density produced by FI, for the same $\Lambda$ value labeling the contiguous dotted line in the UFO region.
}
\label{k2n2relic}
\end{figure}

For example, from Eq.~(\ref{Eq:ratiok2n2})
we have, for $k=2$ and $n=2$ and $m_\chi<\Trh$,
\beq
\Omega_\chi^0h^2|_{k=2,n=2}^{m_\chi<\Trh}\simeq 0.12 \left(\!\frac{\Trh}{1000~\rm GeV}\!\right)^3\!\left(\!\frac{10^{7}\rm GeV}{\Lambda}\!\right)^4 \!
\frac{m_\chi}{29~\rm keV}
\,,
\label{Eq:omegak2n2mlesstrh}
\eeq
for a real scalar DM candidate ($g_\chi=1$) and $g_{RH}=106.75$. We show in Fig.~\ref{k2n2relic}
the allowed region in the parameter space ($m_\chi$, $\Trh$),
for different values of $\Lambda$.

For $m_\chi>\Trh$, we can again consider the case of $k=2$, $n=2$, and take Eq.~(\ref{Eq:ratiomchik2n2}) in Eq.~(\ref{Omega}),
which gives
\begin{align}
\Omega_\chi^0h^2|_{k=2,n=2}^{m_\chi > \Trh} & \simeq
0.12\left(\frac{\Trh}{10^3~\rm GeV}\right)^7\left(\frac{2.4 \times 10^7\rm GeV}{\Lambda}\right)^4 \nonumber \\ & \times \left(\frac{10^5~\rm GeV}{m_\chi}\right)^3\,,
\label{Eq:omegaIRk2n2}
\end{align}
This is effectively the behavior we observe in the region $m_\chi>\Trh$ 
of Fig.~\ref{k2n2relic}.

\begin{figure}[!ht]
\centering
\vskip .2in
\includegraphics[width=3.3in]{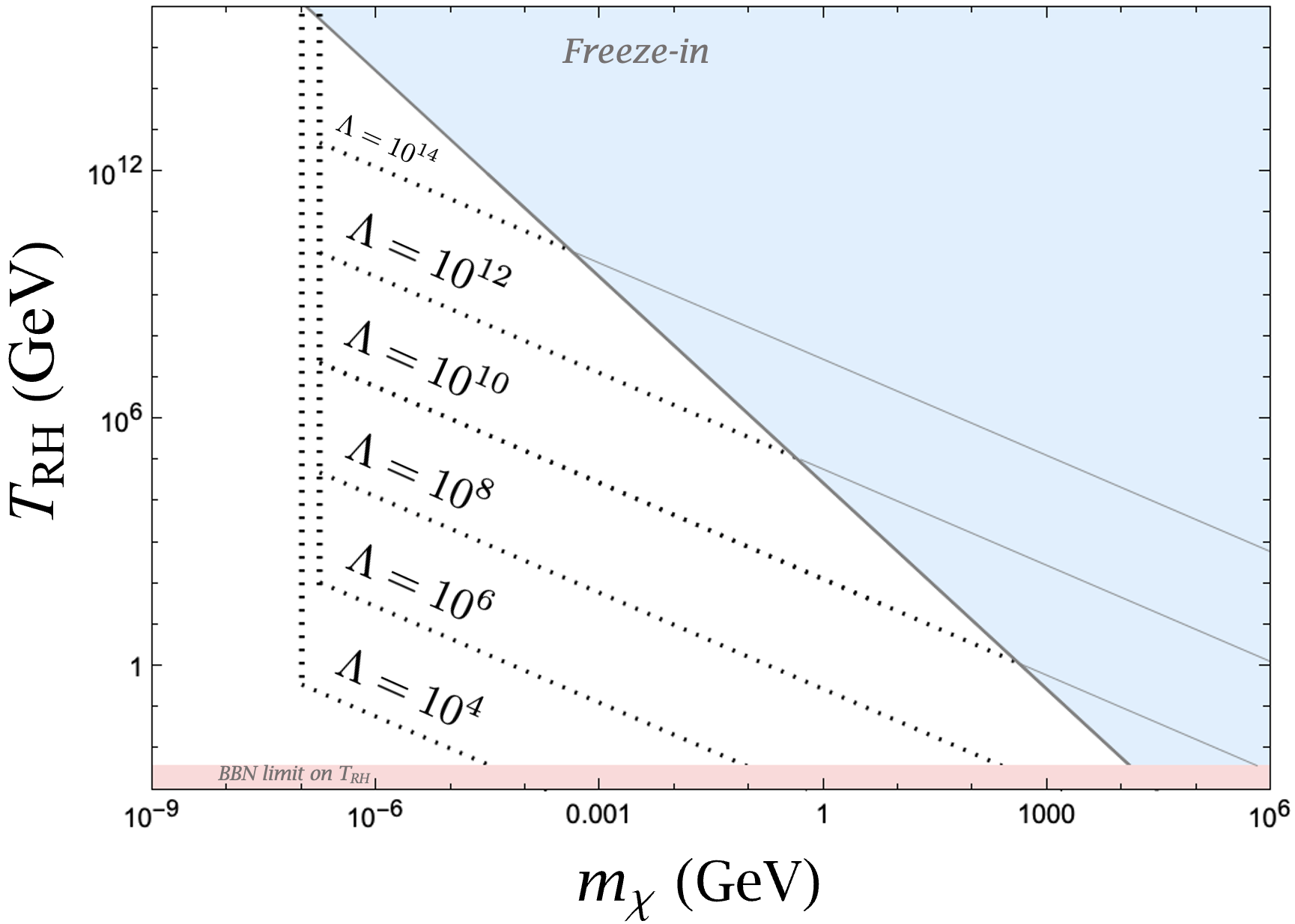}
\caption{\em \small $m_{\chi}$ vs. $T_{\rm RH}$ plane consistent with $\Omega_{\chi} h^2=0.12$ for $k=4$ and $n=2$ with $g=1$, so that $\Lambda=M$. The solid dark gray line corresponds to $\Tfo = M$, and determines the boundary between FI and FO.  The white region corresponds to the correct relic density produced by UFO for each $\Lambda$ listed. The thin gray lines in the blue region correspond to the correct relic density produced by FI, for the same $\Lambda$ value labeling the contiguous dotted line in the UFO region. The vertical dotted lines correspond to UFO after reheating, where the DM mass consistent with the $\Omega_{\chi} h^2=0.12$ is fixed by $g^{*}_{\rm FO}$ alone (see Eq.~ (\ref{Oh2rfo})).
}
\label{k4n2relic}
\end{figure}

In the case of UV production after UFO, we have already remarked that the value
of $\Yrh$ does not depend upon the relative magnitude between 
$m_\chi$ and $\Trh$, because the production occurs 
much before $\am$. We then have only one regime. 
Combining Eqs.~(\ref{Eq:ratio1}) 
and (\ref{Omega}), we deduce
for $k=4$ and $n=2$, and a real scalar $\chi$
\beq
\Omega_\chi^0h^2|_{k=4,n=2}\simeq 0.12 \!
\left(\!\frac{10^{10}\rm GeV}{\Lambda}\!\right)^\frac{12}{7}\!\left(\!\frac{\Trh}{100~\rm GeV}\!\right)^\frac97 \! \frac{m_\chi}{\!1.5~\rm GeV\!}
\label{Eq:omegak4n2}
\eeq
which is depicted in Fig.~\ref{k4n2relic}. Because the production is UV dominated in this case, we do not need to distinguish the regimes $m_\chi>\Trh$ and $m_\chi<\Trh$, in contrast to Fig.~\ref{k2n2relic}.

The behavior seen in Fig.~\ref{k4n2relic} and Eq.~(\ref{Eq:omegak4n2}) can be understood as follows. 
We recall that the cross section with $n=2$ can be interpreted as the exchange of a massive mediator (short range interaction), 
with $\Lambda \sim M$, see Eq.~(\ref{Eq:lambdamu}). For a fixed value of $\Omega_\chi$ and fixed value of $\Lambda$, we see that for large
reheating temperatures, freeze-out happens during the radiation
dominated era, and the constraints on $m_\chi$ are independent of $\Lambda$.
Its only dependence on $\Trh$ comes from the number of 
degrees of freedom which decouple from the bath from $\Tfo$ through $T_0$ today, see Eq.~(\ref{Oh2rfo}). This accounts for the quasi\footnote{``Quasi" in the sense that the number of decoupled 
degrees of freedom depends on the value of $\Trh$.}-vertical line at $m_\chi \sim 100$~eV in Fig.~\ref{k4n2relic}. However, when $\Trh\lesssim \Tfo$, the effect of the continuous 
production of radiation (but not dark matter) during reheating,
for $T < \Tfo$,
begins to affect the ratio between $n_\chi$ and $n_{\rm eq}$, (cf. the lower panel of Fig.~\ref{comovingNumDens2}).
From Eq.~(\ref{Eq:tfokl7}) and setting $\Tfo = \Trh$, we have $\Trh \simeq 286\left(\Lambda / 10^6~\rm GeV\right)^\frac43$ GeV which gives us the point where (for fixed relic density) the quasi-vertical line ought to begin sloping down to higher masses, $m_\chi$ as seen in Fig.~\ref{k4n2relic}. In Fig.~\ref{k4n2relic}, this junction appears at $T_{\rm RH}\approx 120$~GeV, rather than at $286$~GeV. The reason for this is because the sloped lines in Fig.~\ref{k4n2relic} use our analytic result for UV production after UFO given in Eq.~(\ref{Eq:omegak4n2}), which uses the UV limit of integration in Eq.~(\ref{generalYa}). This analytic result holds very well for $\Tfo \gg \Trh$, but as $\Tfo$ approaches $\Trh$, the other limit of integration in Eq.~(\ref{generalYa}) becomes relevant such that when $\Trh=\Tfo$, the estimate in Eq.~(\ref{Eq:omegak4n2}) will be 3-fold greater than the true result. This factor of 3 gets raised to the $\frac{7}{9}$ power when determining $\Trh$ as a function of $m_{\chi}$, such that we find the expected result for the junction in Fig.~\ref{k4n2relic} to be $\Trh=286/\left(3^ \frac{7}{9}\right) \approx 120$~GeV as observed. For lower values of $\Trh$ past the junction in Fig.~(\ref{k4n2relic}), the redshift between $\Tfo$ and $\Trh$ increases, which has the effect of reducing the ratio $n_\chi/n_{\rm eq}$. To compensate for this, 
one needs a higher mass $m_\chi$, following $m_\chi \propto \Trh^{-\frac97}$, as can be seen from Eq.~(\ref{Eq:omegak4n2}) for fixed $\Lambda$. 

However, for a given value of $\Lambda$, one cannot continue to decrease $\Trh$ indefinitely since it leads to larger values of $\Tfo$. At some point $\Trh$ becomes too low and $\Tfo > \Lambda$ preventing $\chi$ from entering into equilibrium. From Eq.~(\ref{Eq:tfokl7})
\beq
\Lambda^\frac57 \lesssim 0.18 M_P^\frac37 \Trh^\frac27\,.
\eeq
Combining this limit with Eq.~(\ref{Eq:omegak4n2}), we obtain the upper limit on $m_\chi$ to ensure 
thermal equilibrium before freeze-out
\beq
m_\chi \lesssim 0.23 ~\left(\frac{10^6~\rm GeV}{\Trh}\right)^\frac35 ~{\rm GeV}\,,
\label{Eq:fi1}
\eeq
as seen in Fig.~\ref{k4n2relic}, 
where the blue-shaded part region represents the freeze-in parameter space described in \cite{HMO}.

Returning to the case with $k=2$ and $n=2$, the out-of-equilibrium production is IR dominated, and the condition to ensure thermal 
equilibrium is still $\Tfo \lesssim \Lambda$. 
If $m_\chi < \Trh$, combining Eqs~(\ref{Eq:tfok2n2}) and (\ref{Eq:omegak2n2mlesstrh})
gives the condition to reach thermal equilibrium as
\beq
m_\chi\lesssim 6.6 \left(\frac{1~\rm TeV}{\Trh}\right)^\frac13~{\rm MeV}\,,
\label{Eq:fi2}
\eeq
as seen in the white region on the left portion of Fig.~\ref{k2n2relic} corresponding to the region $m_\chi<\Trh$. For $m_\chi>\Trh$, combining Eqs.~(\ref{Eq:tfok2n2}) and (\ref{Eq:omegaIRk2n2}), from the conditions $\Tfo \gtrsim m_\chi$ and $\Tfo \lesssim \Lambda$, we obtain 
\beq
870\left(\frac{\Trh}{10 \text{ GeV}}\right)^{\frac{5}{4}}~{\rm GeV} \gtrsim  m_\chi\gtrsim 68~\left(\frac{\Trh}{10~\rm GeV}\right)^\frac{13}{9}~{\rm GeV}\,,
\label{Eq:fi3}
\eeq
which is also what we observe in the region $m_\chi \gtrsim \Trh$ of the Fig.~\ref{k2n2relic}.

To conclude this section, the standard result for relativistic FO limiting the 
mass $m_\chi$ to $\lesssim 100$ eV (see Eq.~(\ref{Oh2rfo})), is {\it only valid in the context of 
instantaneous reheating}. 
However, the decoupling of dark matter {\it during} the 
reheating process significantly relaxes this constraint which 
directly linked the dark matter number density and radiation density.
The main reason for this relaxation comes from the 
fact that the efficient production of radiation continues between $\Tfo$ 
and $\Trh$ due to inflaton decay, while the out-of-equilibrium production of DM is much less efficient, reducing the ratio $n_\chi/n_{\rm eq}$ and allowing much larger masses for $\chi$. This is the case for UV UFO and IR UFO when $m_\chi < \Trh$.

When $m_\chi>\Trh$, the scenario is very different. Indeed, 
the relative increase of the radiation density over the dark matter density during reheating still occurs, and is 
even more efficient when $m_\chi \gtrsim T$. 
For $m_\chi \gtrsim  T$, the thermal bath is no longer able to produce DM, and $n_\chi \propto a^{-3}$.
Relative to the case with $m_\chi<\Trh$, there is additional dilution between $\am$ and $\arh$. To compensate for this (that is to keep $\Omega_\chi h^2$ constant) requires increasing $\Trh$ as $m_\chi$ is increased. To visualize this consider the vertical lines in Fig.~\ref{comovingNumDens2}.
Increasing $m_\chi$ will cause the line $\am/\aend$ to move to the left. Then to keep the ratio of $Y_\chi/Y_{\rm eq}$ fixed, we need to also move the line $\arh/\aend$ to the left. This has the effect of changing the slope of the  iso-density curves, for a given $\Lambda$, and  $m_\chi \propto  \Trh^\frac73$, for $m_\chi > \Trh$ as one can deduce from Eq.~(\ref{Eq:omegaIRk2n2}) and seen in Fig.~\ref{k2n2relic}.

Once again, we leave the cases with $k > 7$ for the ambitious reader.

\subsection{The coldness of dark matter}

Lastly, we examine the important requirement that DM is
cold at the time of structure formation in order to be consistent with the $\Lambda$CDM model. While DM may freeze out 
ultra-relativistically and is therefore considered ``hot" at the time of 
freeze-out, it may have time to cool significantly during reheating as it redshifts as radiation after freeze-out for $m_{\chi}<\Trh$ (for $m_\chi>\Trh$, the DM 
will automatically be non-relativistic (cold) prior to $T_{\rm RH}$). This is a key difference between UFO during reheating vs. UFO during radiation domination. Indeed, for UFO \textit{after} reheating, the values of $m_\chi$ obtained in Eq.~(\ref{Eq:masslimithot}) are excluded because the DM would become non-relativistic too late for small scale structure to form. This warmness constraint may likewise apply to UFO during reheating if freeze-out occurs too close to the reheating temperature, and the DM mass is very small. It is therefore necessary to determine the regions of parameter space in which DM that froze out ultra-relativistically will be cold by the time of structure formation. In addition, we must calculate the contribution of DM to the effective number of degrees of freedom,  $N_{\rm eff}$, and ensure compatibility with the BBN constraint on $N_{\rm eff}$ at $\sim 1$~MeV. We start by deriving the constraint from $N_{\rm eff}$.

\subsubsection{Dark radiation constraint from $N_{\rm eff}$}

We will again consider the cases of UV and IR production separately, since the temperature dependence of the abundance after freeze-out is distinct. We start with UV dominated production. After freeze-out, the DM is relativistic and the co-moving number density is approximately constant. Therefore, after freeze-out, the DM energy density $\rho' = \alpha'T'^4$ redshifts as radiation and \footnote{Of course for $T < \Tfo$, the temperature of the dark matter is only figurative as the DM is out of equilibrium. This is similar to referring to a 1.9 K neutrino background, though the SM neutrinos have been out of equilibrium since $T \lesssim 1$~MeV. The temperature is only used to describe the shape of the momentum distribution function which still scales as if it were in equilibrium with $T \sim a^{-1}$.} 
\beq
T'=\frac{\afo}{a} ~\Tfo
\, .
\eeq
The relationship between $a$ and $T$ is of course different during reheating compared to radiation domination, and so we must evolve the DM temperature through each epoch, until BBN to determine $T'_{\rm BBN}$. Evolving through reheating, we first find
\beq
\Trh'=\frac{\afo}{\arh} ~ \Tfo 
\label{Eq:trhprime}
\, ,
\eeq
where $T'_{\rm RH}$ refers to the temperature of the frozen out DM at the time when the radiation energy density of the universe and the inflaton energy density are equal, which is of course different from $T_{\rm RH}$. 
Combining Eqs.~(\ref{Eq:rhoR}) and (\ref{Eq:trhprime}), we obtain
for\footnote{For $k>7$, the redshift of the thermal bath during reheating is the same as after reheating, $T\propto a^{-1}$.} $k<7$
\beq
\Trh'=\Tfo\left(\frac{\Trh}{\Tfo}\right)^\frac{2k+4}{3k-3}
\left(\frac{g_{\rm RH}}{g_{\rm FO}}\right)^\frac{k+2}{6k-6}
\,.
\eeq
From $\arh$ to the BBN epoch, the Universe is dominated by SM radiation, 
so that
\bea
&&
T'_{\rm BBN}=\left(\frac{\arh}{a_{\rm BBN}}\right)\Trh'=\left(\frac{g_{\rm BBN}}{g_{\rm RH}}\right)^\frac13 \left(\frac{T_{\rm BBN}}{\Trh}\right)\Trh'
\nonumber
\\
&&
= T_{\rm BBN} \left(\frac{\Trh}{\Tfo}\right)^\frac{7-k}{3k-3}
\left(\frac{g_{\rm BBN}}{g_{\rm RH}}\right)^\frac13\left(\frac{g_{\rm RH}}{g_{\rm FO}}\right)^\frac{k+2}{6k-6}\,.
\eea
For $k=2$ we then have
\beq
T'_{BBN}=\frac{g_{\rm BBN}^\frac13 g_{\rm RH}^\frac13}{g_{\rm FO}^\frac23}\left(\frac{\Trh}{\Tfo}\right)^\frac53~T_{\rm BBN}
\,.
\eeq
Then if $\Tfo = \Trh$, $T'_{\rm BBN} = T_{\rm BBN} (g_{\rm BBN}/g_{\rm RH})^{1/3}$, as expected.

At the epoch of BBN, we must ensure that the BSM contribution to the total energy density in radiation respects the upper limit to $N_{\rm eff}$ given in terms of the number of neutrinos (beyond 3)  \cite{ysof}, 
\beq
\Delta N_{\rm eff} = N_{\rm eff} - 3 < 0.18
\eeq
at 95 \% CL.
We will again take scalar DM for specificity; each real scalar contributes and equivalent of (4/7) of a neutrino and the energy density in $\chi$ 
is further reduced (relative a SM neutrino) by $(T'_{\rm BBN}/T_{\rm BBN})^4$ so that the constraint becomes
 \begin{align}
\frac47 \left(\frac{T'_{\rm BBN}}{T_{\rm BBN}}\right)^4 & = \frac47  \left(\frac{T_{\rm RH}}{\Tfo}\right)^{20/3} \left(\frac{g_{\rm BBN} g_{\rm RH}}{g_{\rm FO}^2}\right)^{4/3} < 0.18 
 \end{align}
or 
\beq
\Trh <  0.84 \left(\frac{g_{\rm FO}^2}{g_{\rm BBN} ~g_{\rm RH}}\right)^\frac15 \Tfo \simeq 1.3~\Tfo
\,,
\eeq
where we took $g_{\rm FO}=g_{\rm RH}=106.75$, and $g_{\rm BBN}=10.75$ in the last equality. This constraint is always satisfied since we have implicitly assumed $\Trh < \Tfo$.
Indeed even if $\Tfo = \Trh$, the constraint is satisfied for 
 $(g_{\rm BBN}/g_{\rm RH}) < 0.42$, making it effectively a constraint on $\Trh$.
 In this case, for $g_{\rm FO} = 427/4$
 the scalar never contributes more than 3\% of a neutrino
 and easily satisfies the BBN constraint. Note that this constraint only applies if $m_\chi < T'_{\rm BBN} < 1$~MeV. We also remind the reader that there is nothing special about scalar DM, as UFO is perfectly compatible with fermionic DM.

\subsubsection{Structure formation constraint}

The free--streaming length of a dark matter candidate which was in 
thermal equilibrium with the standard model bath depends on its mass and on the ratio of
its decoupling temperature $\Tfo$ to the reheating temperature. 
Relativistic (or fast) particles will have long free streaming lengths and erase structure on small scales. More precisely, light and relativistic particles present at the time of structure formation will distort the power spectrum and limits to the (warm) DM mass can be set from the Lyman-$\alpha$ forest power spectrum \cite{Irsic:2017ixq}.

Thus, in order to preserve structure on small scales, the dark matter must not be hot (as would be the case for SM neutrinos with eV masses), and must also not be too warm. 
The warmness constraint derived from the Lyman-$\alpha$ forest \cite{Irsic:2017ixq,Barman:2022qgt} 
requires that the typical velocity of the DM
at the time of structure formation (around 1 eV) should be highly non-relativistic, $v_\chi < 2 \times 10^{-4}$ at $T \simeq 1$~eV. 
Taking $v_\chi = p_\chi/m_\chi$ and $p_\chi \simeq \Tfo(\afo/a)$, we can redshift the DM
momentum down to 1 eV using 
\beq
p_\chi \sim T' \simeq \Tfo  \left( \frac{\afo}{\arh} \right) \left({\frac{\arh}{a}}\right)
\label{pchia}
\eeq
so that for $k=2$
\beq
p_\chi \simeq \Tfo  \left( \frac{\Trh}{\Tfo} \right)^{\frac83} \left( \frac{T}{\Trh} \right)
\eeq
and at 1 eV, we have the constraint that 
 \beq
m_\chi > 5~{\rm keV} \left( \frac{\Trh}{\Tfo} \right)^{\frac53}.
\label{mchilim2}
 \eeq
The general constraint for $k<7$ is
 \beq
m_\chi > 5~{\rm keV} \left( \frac{\Trh}{\Tfo} \right)^{\frac{7-k}{3k-3}}.
\label{mchilimk}
 \eeq

Sub-keV DM masses are therefore tolerated if the DM freezes-out well before the end of reheating. For instance, if $T_{\rm RH}=100$~GeV and $T_{\rm FO}=10^5$~GeV and $k=2$, we find that even sub-eV DM particles would be cold by the time of structure formation.

For the case of IR production, DM production continues at temperatures below $\Tfo$. Therefore the initial momenta, $p_\chi$, will
not be subject to redshift from $\Tfo$ as in Eq.~(\ref{pchia}). Instead, even though $\chi$
is not produced in equilibrium at $T < \Tfo$, its momentum will (from freeze-in) production will be characteristic of the temperature of the SM bath, i.e. $T$, so that from the limit on $m_\chi$, can still be obtained from Eq.~(\ref{mchilim2}) or (\ref{mchilimk}) by setting $\Tfo = \Trh$. Namely, $m_\chi \gtrsim 5$~keV, independent of $\Trh$ or $k$. 

\section{Conclusion}
\label{sec:concl}

The dark matter relic density is usually ascribed to one of three mechanisms:
1) Non-relativistic freeze-out, as in the case of a WIMP; 2) Freeze-in, which may occur for very weakly interacting particles (either through weak couplings or heavy mediators) collectively called FIMPs; or, less commonly mentioned, 3) Relativistic freeze out, as in the case of weakly interacting eV DM such as a light neutrino. However, when confronted with experiment, none of these are truly satisfactory. WIMP-like constructions are continually constrained by direct detection experiments \cite{LZ,PandaX:2024qfu,XENON:2025vwd}, which limit the scattering cross section between the dark sector and the Standard Model to roughly 10 orders of magnitude below a typical 
weak cross section. This puts strong constraints on WIMP candidates such as the lightest neutralino \cite{eos}. While the FIMP paradigm is compatible with many DM candidates, most FIMP candidates such as the gravitino \cite{grav} provide very little hope for detection due to extremely small couplings. Furthermore, in most FIMP scenarios studied in the literature, it is
implicitly {\it assumed} that the initial DM abundance is 0. This then lacks the advantage of the WIMP scenario where initial conditions are erased by entering into thermal equilibrium.
Lastly, the classical\footnote{``Classical" in the sense of freeze-out during radiation domination.} relativistic freeze-out scenario implies a tight limit on the DM mass (see Eq.~(\ref{Oh2rfo})) which is in fact excluded by large-scale structure constraints as well as Lyman-$\alpha$ data. We have demonstrated in this report that UFO during reheating is a compelling alternative to these classical scenarios, and may be able to overcome all of the aforementioned challenges.

As we have shown in this work, and as others have previously shown in other contexts, if DM is brought into thermal equilibrium and freezes out {\it during} the reheating phase, the comoving number density of radiation  continues to increase due to inflaton decay, even as the DM density decreases due to the expansion of the Universe.  Thus the relative amount of DM to radiation decreases until reheating is completed and the radiation density is $\sim \Trh^3$. The ratio $n_\chi/n_R \sim \Tfo^3/\Trh^3$ (for UV UFO or WIMP-like FO) can then easily reach $\sim 10^{-8}$, depending on the inflaton width which determines $\Trh$. This phenomenon is well depicted for UV UFO in the bottom panel of Fig.~\ref{comovingNumDens}, which corresponds to an inflaton potential $V(\phi)\sim\phi^4$ (about its minimum)
and a $\chi-$radiation interaction cross section,  $\langle \sigma v \rangle \sim \frac{T^2}{M^4}$, for a mediator mass $M$ at $T \ll M$. 

This extra dilution of $n_{\chi}$ relative to $n_{R}$ due to freeze-out during the reheating period has been identified and studied previously for WIMP-like non-relativistic FO \cite{Silva-Malpartida:2024emu,Mondal:2025awq,ZprimeReheating,ThermalDMBernal}. However, (ultra)relativistic freeze-out has not previously been identified as a viable mechanism for DM production, as it has often been assumed that freezing out (ultra)relativistically will not yield a cold relic \cite{ThermalDMBernal}. However, as we have shown here, there is a vast parameter space associated with DM that freezes out ultra-relativistically, becomes cold before structure formation, and accounts for the entire observed relic density. This UFO mechanism is especially relevant and in fact unavoidable for short-range interactions, for example, those involving a heavy vector or scalar mediator. The UFO scenario also resolves the question of initial conditions, as in the case of a WIMP, because the DM enters into thermal equilibrium with the Standard Model before freezing-out, erasing any pre-history of dark matter production after or during inflation.

In this work, we first systematically derived the conditions for which UFO is possible. In particular we derived the form of the cross-section (its dependence on temperature) which
allows the DM to come into equilibrium and freeze-out ultra-relativistically before reheating is complete. We also derived the ultra-relativistic freeze-out temperature and the constraints on the reheating temperature such that the UFO scenario is possible. In doing so, we delineated reasonably accurate estimates of the UFO/FI boundary, as well as the UFO/WIMP boundary. The UFO/FI boundary lies approximately in the regions of parameter space where $\Tfo=M$ (see dark gray lines in Figs.~\ref{k2n2relic} and \ref{k4n2relic}), where $M$ is the mass of a heavy mediator. The UFO/WIMP boundary lies in the regions of parameter space where $\Tfo \approx m_{\chi}$. However, it is worth noting that the more precise boundaries will depend on the exact form of the interaction. For instance, the width of the mediator will impact the UFO/FI boundary, while any lower order terms in $T$ in the thermally averaged cross section will impact the UFO/WIMP boundary. We leave these more detailed considerations for future work. These boundaries are quite important, since short range interactions will typically exhibit all three of these behaviors (WIMP-like FO, UFO, and FI), depending on the specific combination of $\Lambda$, $\Trh$, and $m_{\chi}$. Moreover, it is worth noting that while the WIMP to FIMP transition has previously been studied during the reheating era \cite{Silva-Malpartida:2024emu, ThermalDMBernal, ZprimeReheating}, UFO has been neglected entirely, likely because it is impossible for some interactions (e.g. a contact interaction between DM and SM scalars) but possible for others, along with the aforementioned misconception that UFO will necessarily produce warm DM.

There are several important differences between the UFO scenario and WIMP-like non-relativistic freeze-out. In non-relativistic freeze-out, DM production quickly ceases for $T<m_\chi$; and after freeze-out $n_\chi a^3$ is constant, providing the final relic density of dark matter. For UFO, significant DM production may occur after freeze-out, similar to the freeze-in mechanism. Another important difference is the allowed values of the couplings. Standard couplings for WIMP-like FO are on the order of the electroweak scale. When WIMP-like FO occurs during reheating, the allowed couplings are reduced to some extent \cite{Silva-Malpartida:2024emu}. However, for UFO, the allowed range of couplings is comparably quite vast. For instance, see Fig.~\ref{k2n2relic}, where we see that for heavy mediator interactions, UFO is compatible with the observed relic abundance for $10^{3} \lesssim \Lambda \lesssim 10^{14}$~GeV! This is a much broader range of couplings than is permitted by WIMP-like freeze-out, either during or after reheating, and is more likely to evade experimental constraints from direct detection. Finally, WIMP-like FO during reheating typically requires lower reheating temperatures ($\Trh \lesssim 10^{5}$~GeV) to produce the correct relic density \cite{ThermalDMBernal}. In contrast, UFO can produce $\Omega_{\chi} h^2=0.12$ for $\Trh$ values spanning about 17 order of magnitude (see Fig.~\ref{k2n2relic}).

There are likewise several key differences between UFO and the freeze-in mechanism. While freeze-in assumes zero initial DM abundance, the initial condition for UFO is determined by the abundance at freeze-out. This is especially important when the out-of-equilibrium production is UV dominated, in which case a naive freeze-in calculation would produce very different results from the proper UFO calculation. In contrast, when the out-of-equilibrium production is IR dominated, a naive freeze-in calculation (where one takes the incorrect assumption that the interaction never reached equilibrium) would actually produce essentially the correct result given by UFO, since almost all of the production happens in the IR regime near $\Trh$ or $m_{\chi}$, well after FO. However, perhaps the most important distinction between UFO and freeze-in is the allowed range of couplings. Typical couplings for FIMPs are very feeble, often on the order of $10^{12} \lesssim \Lambda \lesssim 10^{18}$~GeV \cite{HMO}. In contrast, UFO couplings are on the order of $10^{3} \lesssim \Lambda \lesssim 10^{14}$, which makes UFO much more accessible to direct and indirect detection efforts compared to FIMPs.

Next, we note the key distinction between classical relativistic freeze-out during radiation domination and UFO during reheating. In the classical relativistic case where freeze-out occurs after reheating, there is no Boltzmann suppression and $n_\chi a^3$ is constant after freeze-out. In this case, no out-of-equilibrium production can occur after freeze-out since the co-moving number density of the source particles (SM radiation) is constant. This leads to the very tight constraint on the mass of the DM (as in the case of neutrinos), which is in fact excluded due to constraints from structure formation (see Eq.~(\ref{Oh2rfo}) and surrounding discussion). In the UFO scenario described here, there is also never a Boltzmann suppression of the density. However, the tight constraint on the mass is significantly relaxed (the allowed DM mass range can expand by over 10 orders of magnitude) since substantial out-of-equilibrium production and/or dilution may occur from $\Tfo$ to $\Trh$. This allows for a much richer parameter space consistent with the observed relic density, compared to classic relativistic FO. The extra dilution during reheating also allows for light but cold dark matter, unlike the classical case.

We also found that there is a distinction between UV and IR UFO, which further enriches the possibilities for the UFO scenario. For UV UFO, DM production will effectively cease near the freeze-out temperature, which is covered in detail in section \ref{sec:DMden}. Alternatively, if production is IR dominated, thermal production stops at $\Tfo$, but non-thermal production continues as in a FI scenario. However, unlike the FI case, there is no arbitrariness in the initial conditions for the DM density (usually assumed to be 0 in the standard FI case), since DM was previously in thermal equilibrium (as in the standard FO case). The distinction between UV and IR production is determined by the particular interactions between the DM and SM along with the particular reheating model. 

There are some limitations of this study. First, because we conducted a general investigation of UFO by treating multiple types of interaction cross sections and reheating models, we simplified our analysis by taking $\mathcal{O}(1)$ couplings when possible. In general, the regions of parameter space compatible with FI vs. UFO vs. WIMP-like FO are of course sensitive to the couplings. For smaller couplings, both the FI lines and UFO lines for a given $\Lambda$ will move upward in the ($m_{\chi}, T_{\rm RH}$) plane in Figs. \ref{k2n2relic} and \ref{k4n2relic}. Additionally, for the high temperatures reached during reheating, we expect that the couplings will run. We did not treat the running of the couplings here. Next, our analysis here provides model-independent results for $2\rightarrow2$ SM-to-DM processes. As a result, when a particular model is specified, a more detailed analysis using all terms in the relevant Lagrangian will be necessary. Finally, because we were primarily interested in UFO, we parametrized the thermally averaged cross section in the relativistic regime, which neglects lower order terms in $T$, for instance those depending on $m_{\chi}$. This will impact the UFO/WIMP boundary since these terms become relevant as we approach the non-relativistic regime. However, these lower order terms are model-dependent, so we will leave a more detailed analysis of the UFO/WIMP boundary for future work.

Ultimately it is experiment that must determine the nature of DM. The UFO DM scenario greatly enhances the parameter space (in mass and coupling strength) which may be accessible to experimental searches. In fact, the UFO scenario requires interaction strengths that lie precisely between the typical WIMP and FIMP regimes. As a result, UFO-associated dark sectors can simultaneously evade the WIMP detection limits and overcome the challenges associated with vanishingly small FIMP couplings.

\section*{Acknowledgements} \label{sec:acknowledgements}
  This project has received support from the European Union's Horizon 2020 research and innovation program under the Marie Sklodowska-Curie grant agreement No 860881-HIDDeN.
  The work of K.A.O.~was supported in part by DOE grant DE-SC0011842  at the University of
Minnesota. Y.M. acknowledges support by Institut Pascal at Université Paris-Saclay during the Paris-Saclay Astroparticle Symposium 2024, with the support of the P2IO Laboratory of Excellence (program “Investissements d’avenir” ANR-11-IDEX-0003-01 Paris-Saclay and ANR- 10-LABX-0038), the P2I axis of the Graduate School of Physics of Université Paris-Saclay, as well as the CNRS IRP UCMN.

\end{document}